\documentclass[12pt,a4paper]{article}

\usepackage{natbib}
\usepackage[T1]{fontenc}
\usepackage{palatino}
\usepackage{graphics}
\usepackage[demo]{graphicx}
\usepackage{booktabs,caption}
\usepackage{threeparttable}
\usepackage{adjustbox}
\usepackage[format=plain, labelsep=period, labelfont=bf, font=footnotesize,singlelinecheck=false,skip=5pt]{caption}
\usepackage{tabularx}
\usepackage{array}
\newcolumntype{H}{>{\setbox0=\hbox\bgroup}c<{\egroup}@{}}
\usepackage{amsthm}
\usepackage{lscape}
\usepackage{appendix}
\usepackage[bookmarks]{hyperref}
\usepackage{subcaption}
\usepackage{epstopdf}
\usepackage{amsmath}

\usepackage{multirow}
\usepackage{setspace}
\usepackage{amssymb}
\usepackage{textcomp}
\usepackage{amsfonts}
\usepackage{soul}
\usepackage{bbm}
\usepackage{multirow}
\usepackage{multicol}
\usepackage{verbatim}
\usepackage{mdwlist}
\usepackage{color,soul}
\usepackage{eurosym}
\usepackage{rotating}
\usepackage{lmodern}
\usepackage{color}
\usepackage{longtable}
\usepackage{etoolbox}

\usepackage{epigraph}
\usepackage{xargs}
\usepackage[left=2cm,right=2cm,bottom=2cm,top=2cm]{geometry}
\usepackage{ifpdf}
\usepackage[colorinlistoftodos,prependcaption,textsize=normalsize]{todonotes}
\makeatletter

\newcommand*{\@rowstyle}{}
\newcommand*{\rowstyle}[1]{
	\gdef\@rowstyle{#1}%
	\@rowstyle\ignorespaces%
}
\newcolumntype{=}{
	>{\gdef\@rowstyle{}}%
}
\newcolumntype{+}{
	>{\@rowstyle}%
}
\makeatother

\title{Young people between education and the labour market during the COVID-19 pandemic in Italy}

\author{Davide Fiaschi\thanks{University of Pisa, Dipartimento di Economia e Management, Via Ridolfi 10, 56124 Pisa (Italy), Phone: +39 050 2216208, Email: davide.fiaschi@unipi.it. }  \and Cristina Tealdi\thanks{Corresponding author. Department of Economics, Heriot-Watt University, EH14 4AS Edinburgh (UK) and IZA Institute of Labor, Phone: +44 0131 4513803, Email: c.tealdi@hw.ac.uk. } \\
}

\date{\today}

\begin{document}
	
	\maketitle
	
	\begin{abstract}
		We analyse the distribution and the flows between different types of employment (self-employment, temporary, and permanent), unemployment, education, and other types of inactivity, with particular focus on the duration of the school-to-work transition (STWT). The aim is to assess the impact of the COVID-19 pandemic in Italy on the careers of individuals aged 15-34. We find that the pandemic worsened an already concerning situation of  higher unemployment and inactivity rates and significantly longer STWT duration compared to other EU countries, particularly for females and residents in the South of Italy. In the midst of the pandemic, individuals aged 20-29 were less in (permanent and temporary) employment and more in the NLFET (Neither in the Labour Force nor in Education or Training) state, particularly females and non Italian citizens. We also provide  evidence of an increased propensity to return to schooling, but most importantly of a substantial prolongation of the STWT duration towards permanent employment, mostly for males and non Italian citizens. Our contribution lies in providing a rigorous estimation and analysis of the impact of COVID-19 on the carriers of young individuals in Italy, which has not yet been explored in the literature.
	\end{abstract}
	
	\noindent \textbf{Keywords}: Labour market flows, transition probabilities, first passage time, school-to-work transition, NLFET. 
	
	\noindent \textbf{JEL Classification}: I2, J13, J21, J24.
	
	\newpage 
	
	\section{Introduction}\label{sec:introduction}
	
	The devastating economic impact of the COVID-19 pandemic has been shown to differ widely among  demographic and socioeconomic groups \citep{chetty2020economic}. Although the issue of young individuals being one of the categories hit the most has been vocally raised in Europe \citep{european2020living} and extensively acknowledged in the literature \citep{lee2021hit, alon2021mancession, bluedorn2021gender, adams2020inequality, blustein2020unemployment}, surprisingly there are few specific quantitative studies on the topic. In this paper, we investigate how the COVID-19 pandemic has affected the careers of young individuals aged 15-34 in Italy. Specifically, we focus on three main issues, which have been widely debated in the literature \citep{carcillo2015neet, hewins2021predicting, pastore2014youth}: i) the duration of the transition from school to work (STWT), ii) the fall into the NLFET (Neither in the Labour Force nor in Education or Training) state; and, iii) the return to schooling. 
	
	Young individuals in Europe tend to take quite a long time to progress from the end of their studies to the attainment of a job  \citep{pastore2014youth}. Although there is large heterogeneity across countries, young individuals in Mediterranean countries perform particularly poorly and Italy generally ranks last in Europe for the time young take to complete their STWT. In 2017, on average, Italian students took 2.88 years to transit to a job which is at least six month long, compared to an European average of 1.47 years; moreover, for individuals with compulsory education, the duration increases to 5.18 years \citep{pastore2021some}. Many reasons have been identified in explaining these alarming statistics. From a macroeconomics perspective, a longer STWT correlates with GDP growth, the spread of temporary employment and the quality of institutions directly affecting the ability of entrants into labour market to smoothly complete the transition, namely, the educational system, the legal labour market arrangements, and  active labour market policies (ALMP) \citep{pastore2014youth, dietrich2016youth, pastore2020stuck, o2012time}.
	
	Sparse evidence suggests the COVID-19 pandemic has exacerbated these issues \citep{ecobservatory2021}, not only because in the first two quarters of 2020 several young individuals lost their jobs or got furloughed, but also because of the severe drop in GDP and the deterioration of the labour market conditions. As a consequence, young individuals may have decided to delay the transition to the labour market by staying longer in schooling, or may have decided to go back to studying. Most worryingly, they might have also got discouraged and fallen into the NLFET state. 
	In particular, there is some evidence that young workers have been impacted by COVID‐19 significantly more compared to older workers \citep{churchill2021covid, nunes2021failing}. Young women, especially those in their 20s, seem to have been particularly exposed to the economic fallout. There are also studies showing that a large number of people between the age of 15 and 24 have dropped out of the labour market entirely \citep{jackson2020coming, mayhew2020covid}. However, some studies also show that although younger workers lost more jobs initially, no systematic difference remains in the pandemic’s employment impact across age groups later on \citep{lee2021hit}. For instance, in the USA by November 2020, the unemployment effect of the pandemic was fairly similar across all age groups, except for the youngest group’s non-participation rate which had still not fully recovered.
	For the case of Italy, there is some evidence \citep{bancaditalia2021, quaranta2020first} that young workers might be the ones most affected by the pandemic for a number of reasons: (i) the share of young people in the NLFET state was already high before the pandemic, (ii) young people are mainly hired on temporary contracts, (iii) young workers are over-represented in sectors that were shut down during the lockdown and (iv) young workers are mainly employed in sectors that will be structurally changed by the distancing measure.
	
	However, to the best of our knowledge a rigorous estimation and analysis of the impact of COVID-19 on the carriers of young individuals in Italy is missing. In this paper we fill this gap using longitudinal quarterly labour force data for the period 2013-2020, which was the first country in Europe to be hit by the COVID-19 pandemic and the first to implement a national lockdown in the beginning of March 2020.
	We illustrate how the actual shares of individuals in the population aged 15-34 have changed across seven states (self employment, permanent employment, temporary employment, unemployment, education, furlough scheme, and NLFET) before and during the COVID-19 pandemic. Following the literature which provides evidence of longer STWT for low-educated, immigrant, females and resident in the South of Italy \citep{pastore2014youth}, we also calculate the shares for specific groups of individuals identified on the base of gender, (non) citizenship, and geographical area.
	We then estimate the transition probabilities between these seven states, for which we also compute the first passage time (FPT) and the expected first passage time (EFPT). The EFPT, when calculated for the transition from  education to different employment states (temporary and permanent), represents our estimate of the STWT duration (Appendix \ref{app:firstpassagetime}).
	
	 	
	The paper is organized as follows. Section \ref{sec:dataset} illustrates the data. Section \ref{sec:prePostDecretoDignità} discusses the Italian evidence on the duration of the transition from school to work, the fall into the NLFET state and the return to schooling before and during the pandemic. Section \ref{sec:concludingRemarks} contains some concluding remarks.

	\section{Data \label{sec:dataset}}
	
	We use Italian quarterly longitudinal labour force data as provided by the Italian Institute of Statistics (ISTAT) for the period 2013 (quarter I) to 2020 (quarter III).\footnote{Data for the period 2013 (quarter I) to 2020 (quarter III) are available upon request at: https://www.istat.it/it/archivio/185540. Dataset and codes are available at \url{https://people.unipi.it/davide_fiaschi/ricerca/}.} The Italian Labour Force Survey (LFS) follows a simple rotating sample design, where households participate for two consecutive quarters, exit for the following two quarters, and come back in the sample for other two consecutive quarters. As a result, 50\% of the households, interviewed in a quarter, are re-interviewed after three months, 50\% after twelve months, 25\% after nine and fifteen months. This rotation scheme allows to obtain 3 months longitudinal data, which include almost 50\% of the original sample.
	On average approximately 70.000 individuals are interviewed each quarter, of which 45.000 are part of the working age population. 
	
	Per each interviewee we observe a large number of individual and labour market characteristics at the time of the interview and three months before. We then compute the labour market flows by calculating the quarter-on-quarter transitions made by individuals between seven different states: permanent employment, temporary employment, self-employment, unemployment, education, furlough scheme and NLFET. The drawback of these data is the point-in-time measurement of the individuals' state, which fails to capture the transitions within the period (quarter).
	
	\section{Young individuals before and during the COVID-19 pandemic} \label{sec:prePostDecretoDignità}
	
	In Italy the COVID-19 epidemic started spreading exponentially in the beginning of March 2020 and a full lockdown was implemented on March 10, 2020. The first impact of the pandemic in quarter I of 2020, and a much more pronounced impact in the following quarter II, when the lockdown was fully in place.The quarter III of 2020 should reflect the relaxing of lockdown regime and incidence of pandemic.
	We consider four categories of individuals \citep{pastore2014youth}: teens (age 15-19), early young (age 20-24), late young (age 25-29) and pre-adults (age 30-34) and we mainly focus on the two (early and late) young categories, taking teens and pre-adults  as reference. 
	In Section \ref{sec:labourShares}, for each age category we look at how the shares of individuals in the seven states have evolved in the period of observation 2013-2020; in Section \ref{sec:transitionProbabilities} we compute transition probabilities between states; and, finally, in Section \ref{sec:FPT} we compute FPT and EFPT for the transitions from education to permanent and temporary employment.
	
	\subsection{The evolution of the shares of individuals across different states}\label{sec:labourShares}
	
\begin{figure}[htbp]
		\caption{Shares of individuals aged 15-34 in different states by age categories.}
		\label{fig:shares}
		\vspace{0.3cm}
			\begin{subfigure}[b]{0.32\textwidth}
		\centering
		\includegraphics[width=1\linewidth]{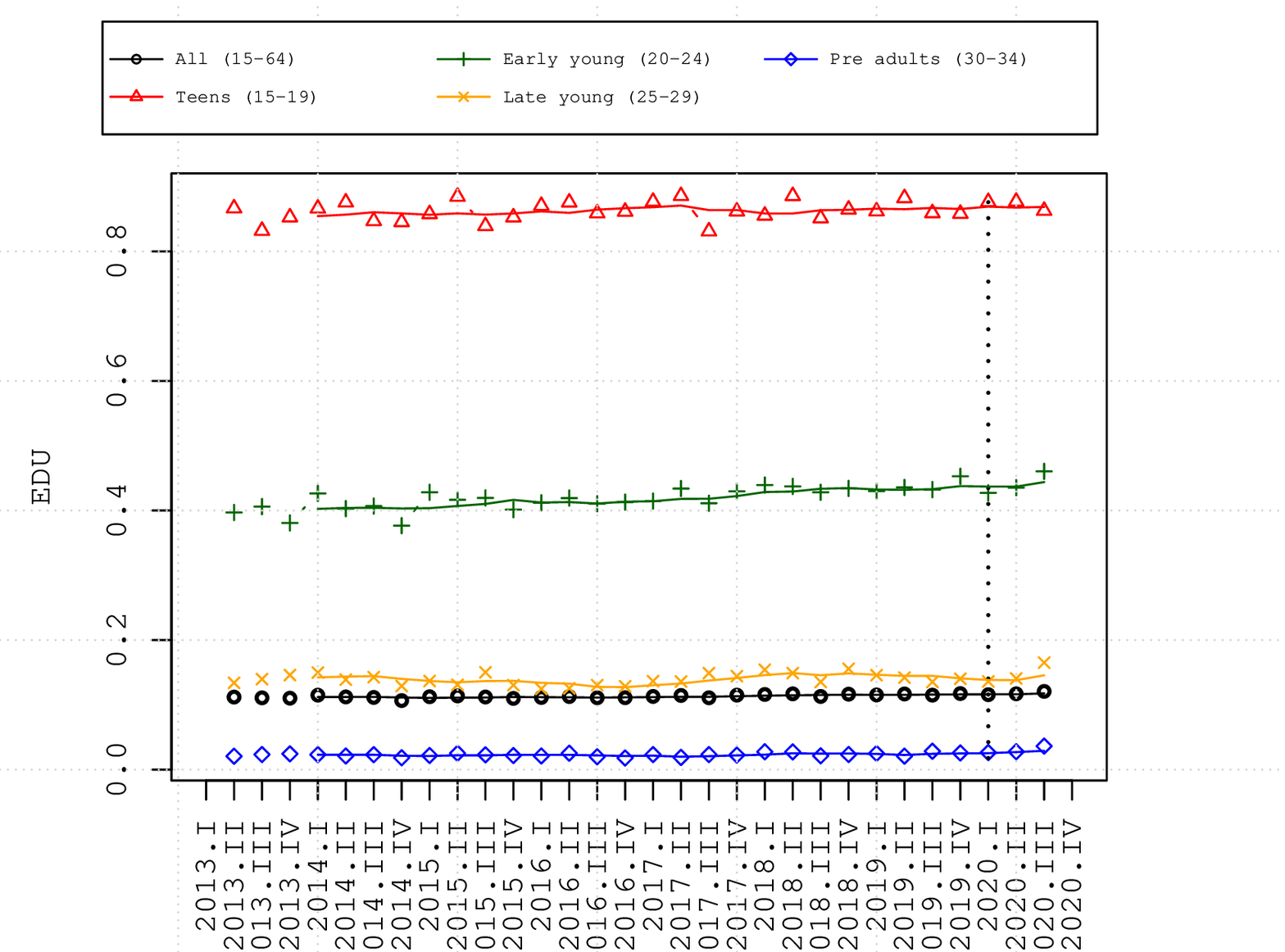}
			\caption{Education.}
		\label{fig:actualMassEDU}
		\vspace{0.2cm}
			\end{subfigure}
				\begin{subfigure}[b]{0.32\textwidth}
			\centering
			\includegraphics[width=1\linewidth]{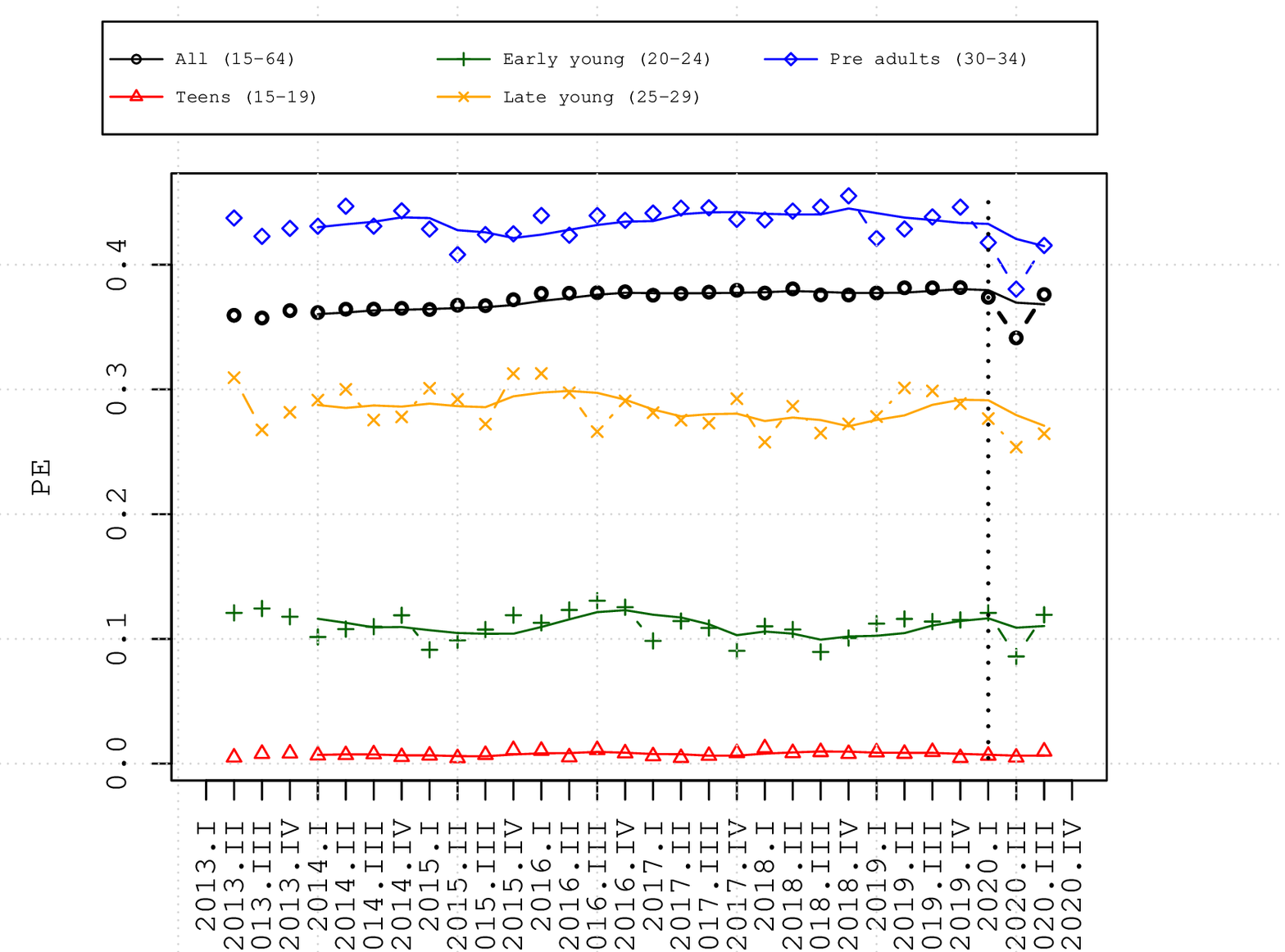}
				\caption{Permanent employment.}
			\label{fig:actualMassPE}
		\end{subfigure}
			\begin{subfigure}[b]{0.32\textwidth}
		\centering
		\includegraphics[width=1\linewidth]{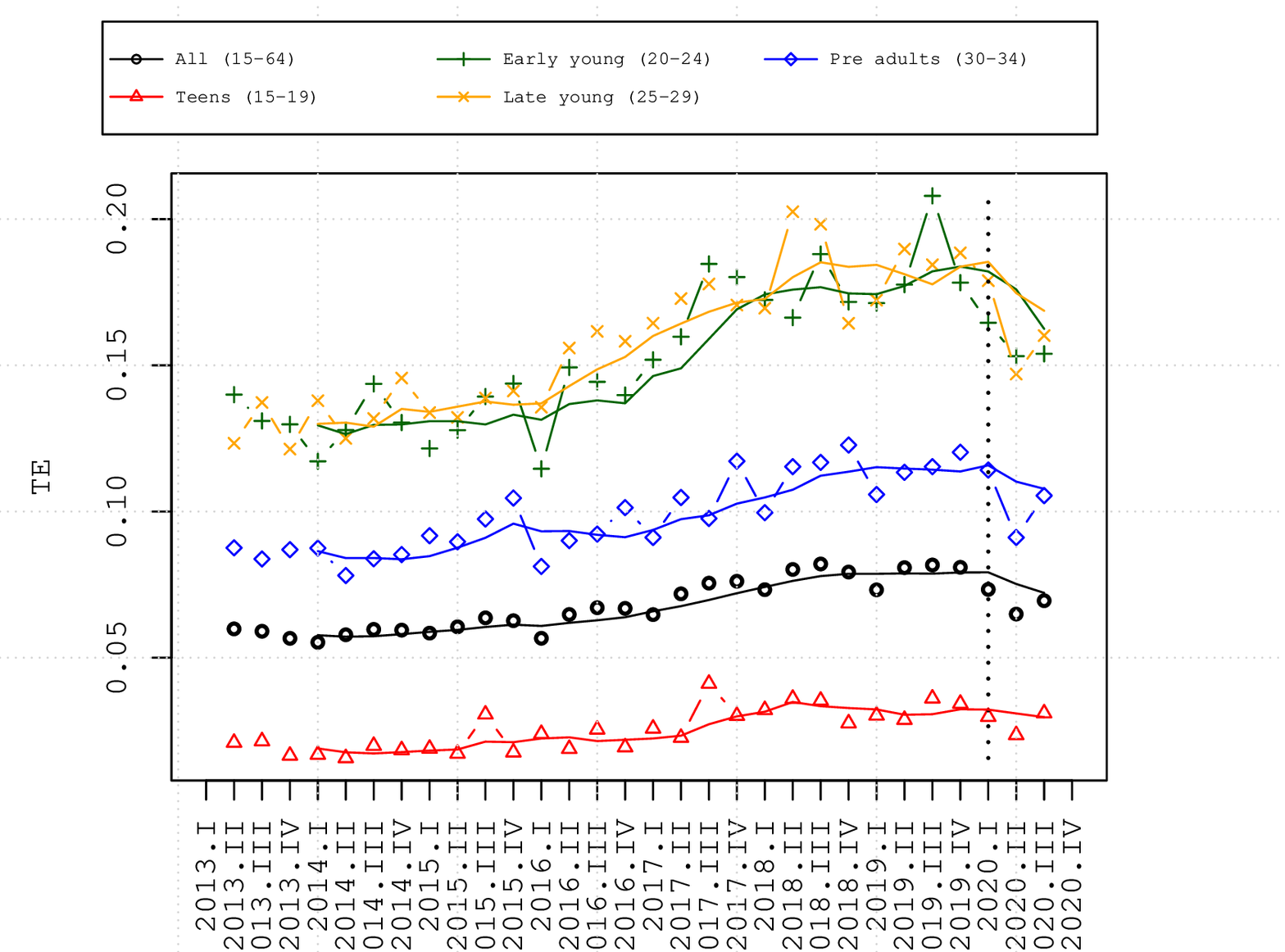}
			\caption{Temporary employment.}
		\label{fig:actualMassTE}
							\vspace{0.2cm}
	\end{subfigure}
		\begin{subfigure}[b]{0.32\textwidth}
	\centering
	\includegraphics[width=1\linewidth]{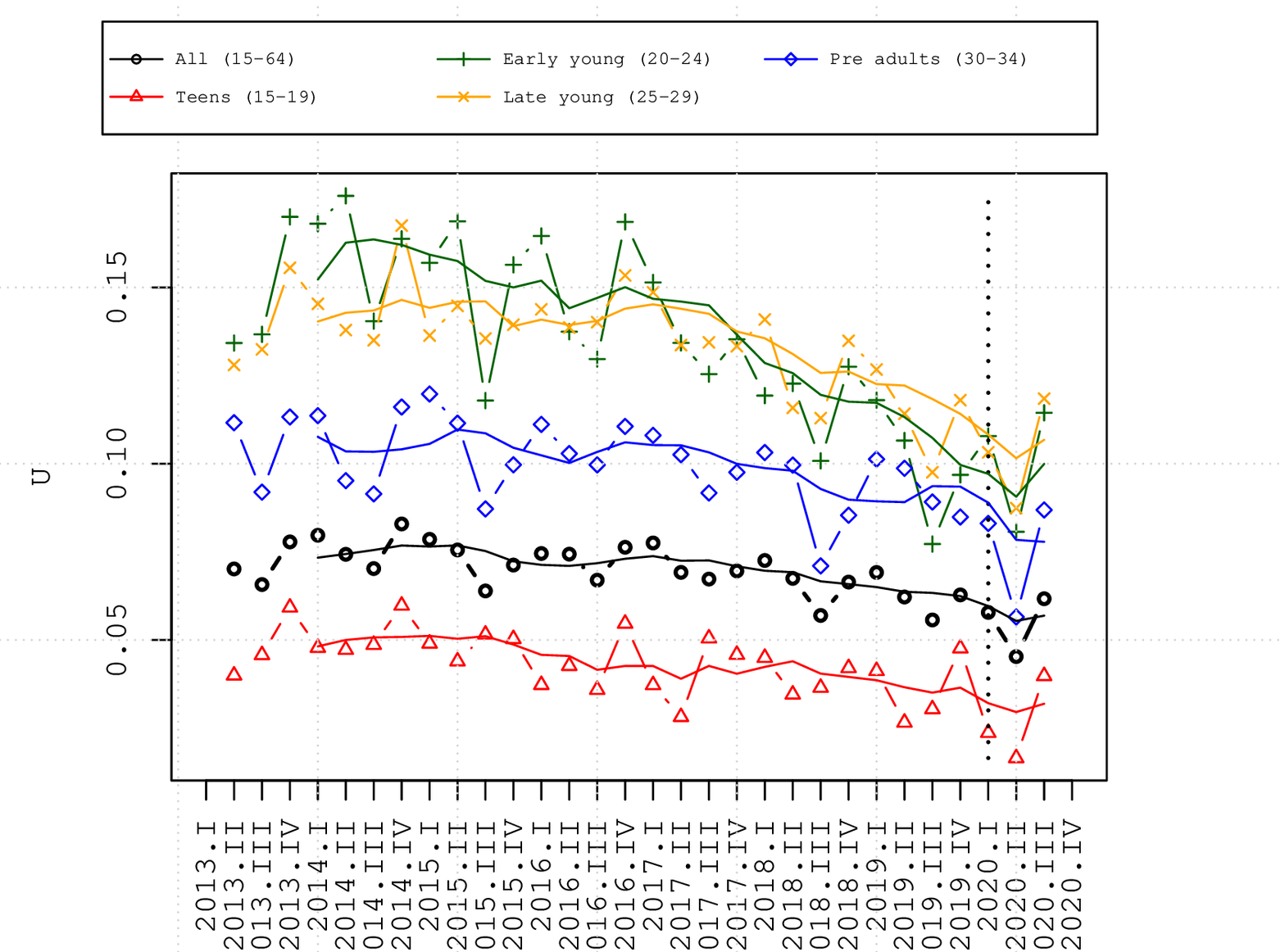}
		\caption{Unemployment.}
	\label{fig:UactualMassU}
\end{subfigure}
		\begin{subfigure}[b]{0.32\textwidth}
	\centering
	\includegraphics[width=1\linewidth]{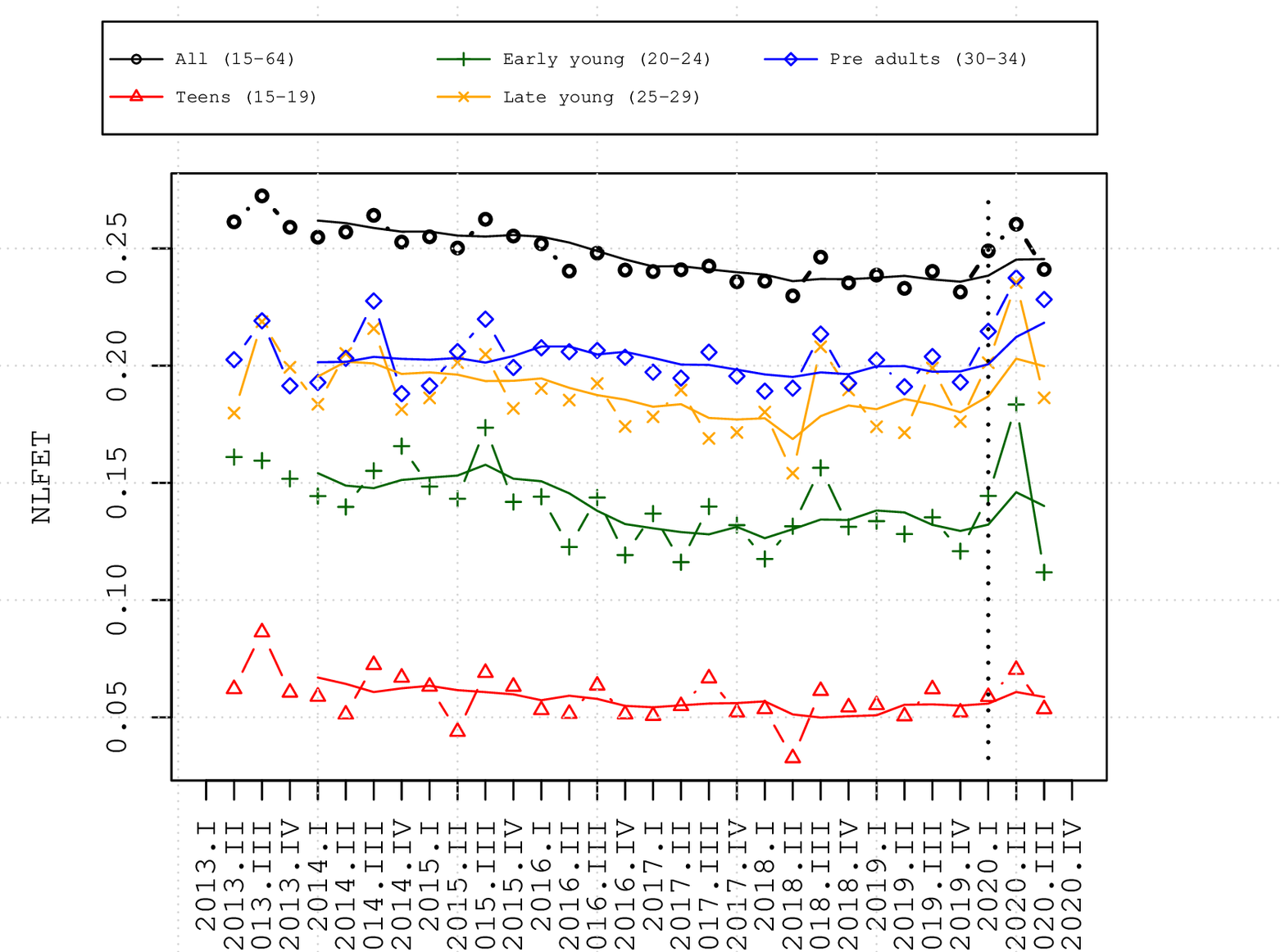}
		\caption{NLFET.}
	\label{fig:actualMassNLFET}
\end{subfigure}
		\begin{subfigure}[b]{0.32\textwidth}
	\centering
	\includegraphics[width=1\linewidth]{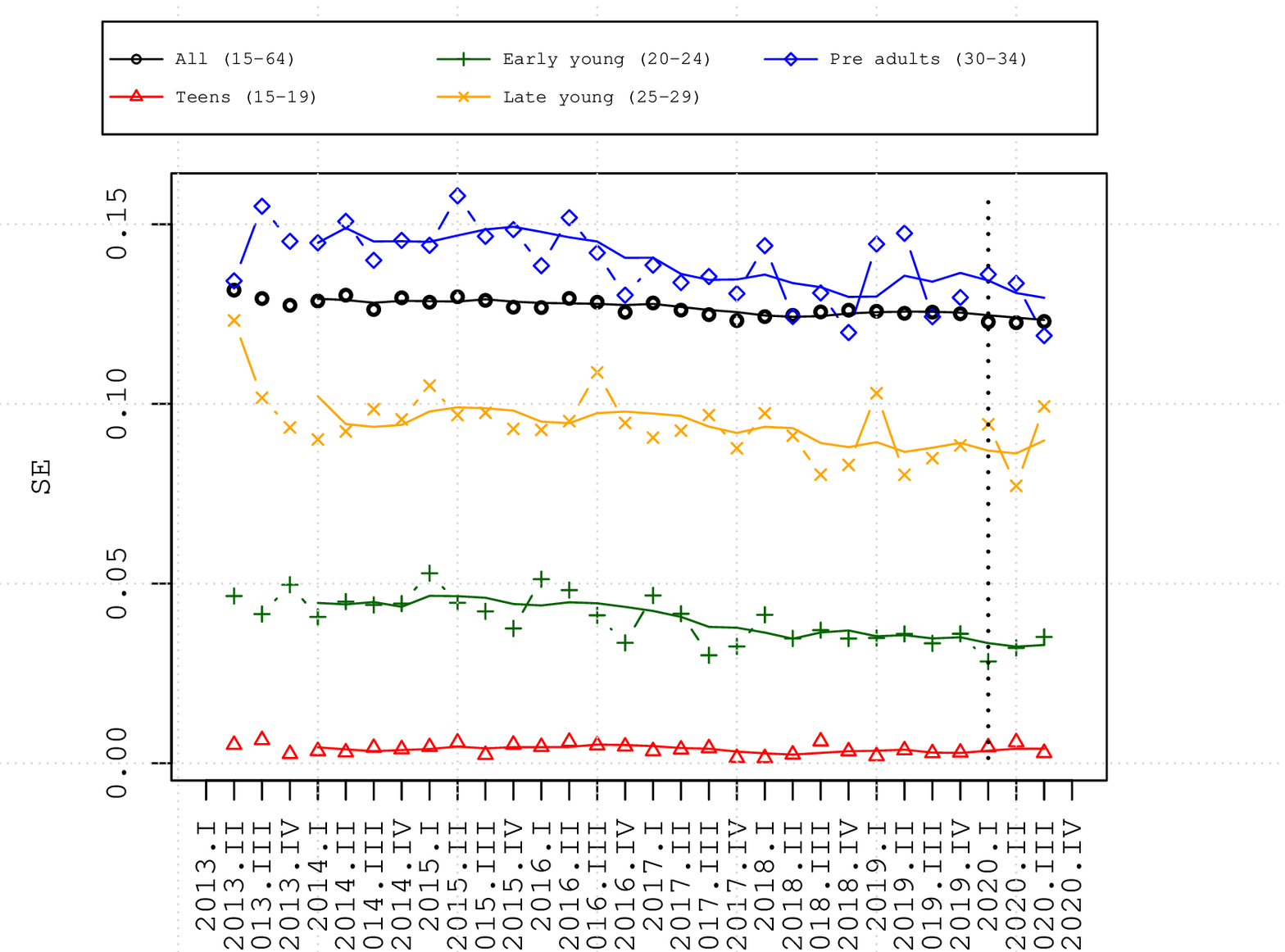}
	\caption{Self-employment.}
	\label{fig:actualMassSE}
\end{subfigure}
\begin{subfigure}[b]{0.32\textwidth}
\centering
\includegraphics[width=1\linewidth]{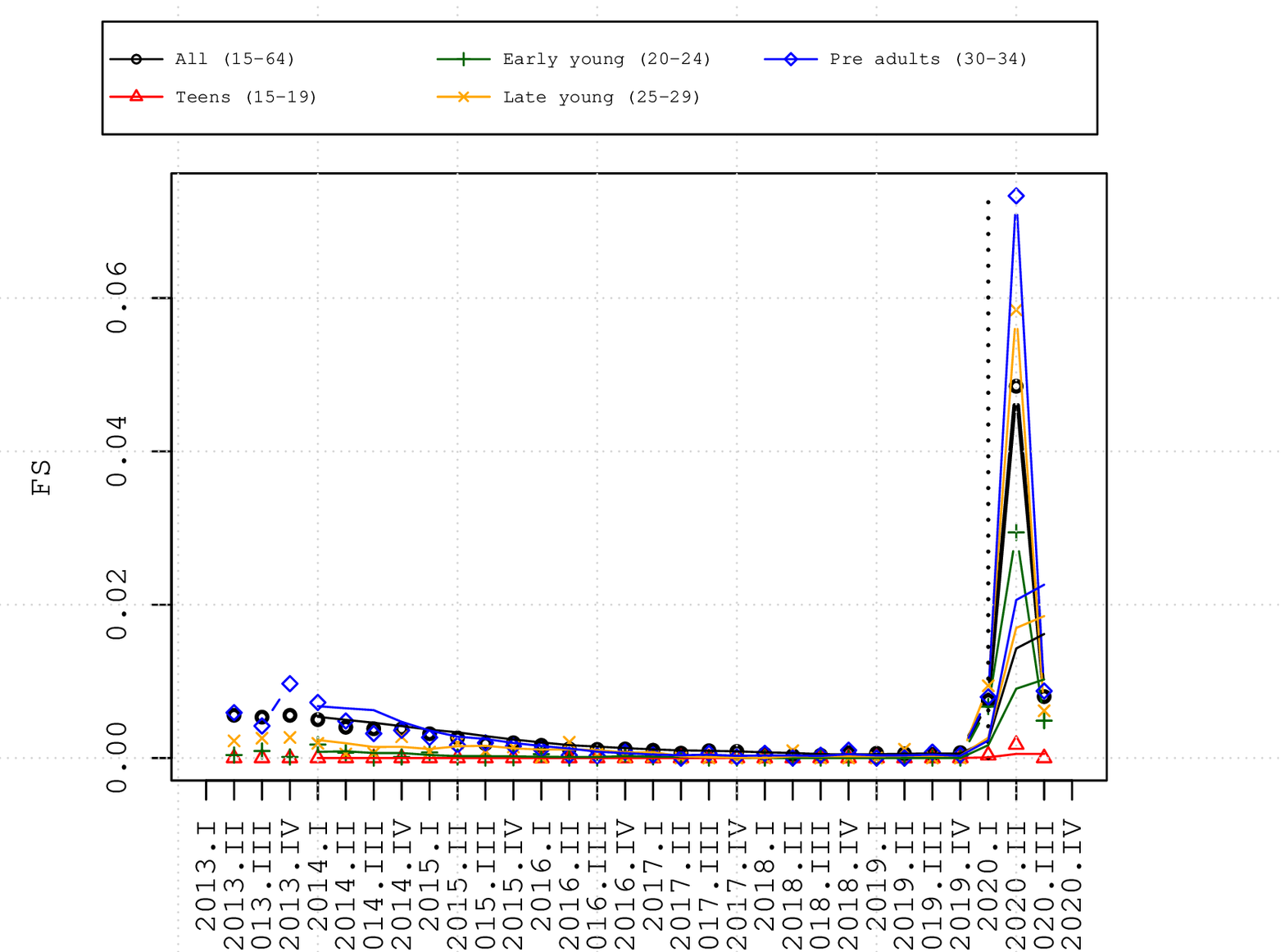}
\caption{Furlough scheme.}
\label{fig:actualMassFS}
\end{subfigure}
\caption*{\scriptsize{\textit{Note}: EDU refers to education, PE to permanent employment, TE to temporary employment, U to unemployment, SE to self-employment and FS to the furlough scheme, NFLET to not in the labour force, education or training. Source: LFS 3-month longitudinal data as provided by the Italian Institute of Statistics (ISTAT).}}
	\end{figure}

	Table \ref{fig:shares} reports the shares of individuals in different states from quarter I of 2013 to quarter III of 2020.
	In the quarters just before the pandemic approximately 43\% of early young were in education, 11\% in permanent employment, 19\% in temporary employment, 4\% in self-employment, 9\% in unemployment and 13\% in the NLFET state. As expected, the share of late young in education was smaller (approximately 14\% versus 43\%), while the share in employment was larger (19\% versus 11\% in permanent employment, 30\% versus 19\% in temporary employment, 8\% versus 4\% in self-employment). The share of late young in unemployment was comparable to the share of early young (10\% approximately), while a larger share of late young was in the NLFET state (19\% versus 13\%). A similar share of pre-adults (approximately 20\%) was also in the NLFET state.
	
The pandemic has lead to a marked raise in the share of both early and late young in education; in quarter III of 2020 these shares increased from 43.5\% to 46\% and from 14.1\% to 16.5\%, respectively. Instead, its impact on the share of permanent employment was strongly negative and uniform across all age categories in quarter II of 2020. However, in  quarter III of 2020 while for teens and early young the permanent employment share bounced back to the pre-pandemic value, for late young and pre-adults the share did not fully recover. Specifically, for late young it declined from 29.1\% pre-pandemic to 26.5\% in quarter III of 2020. Similar patterns are observed for temporary employment: all categories suffered a drop in the temporary employment share in quarter II of 2020, stronger among early and late young, but only teens fully recovered in quarter III of 2020. For all other categories, the bounce back was not strong enough to achieve the pre-pandemic values. For early young the average temporary employment pre-pandemic share was 18.3\% compared to a share of 15.4\% in quarter III of 2020. The unemployment share decreased in quarter II and bounced fully back for all categories in quarter III. 

\begin{figure}[htbp]
		\caption{Shares of individuals aged 15-34 in the NLFET state by individual characteristics (age categories, gender, and citizenships).}
		\label{fig:sharesfemaleNEET}
		\vspace{0.2cm}
		\centering
		\begin{subfigure}[t]{0.32\textwidth}
			\centering
			\includegraphics[width=1\linewidth]{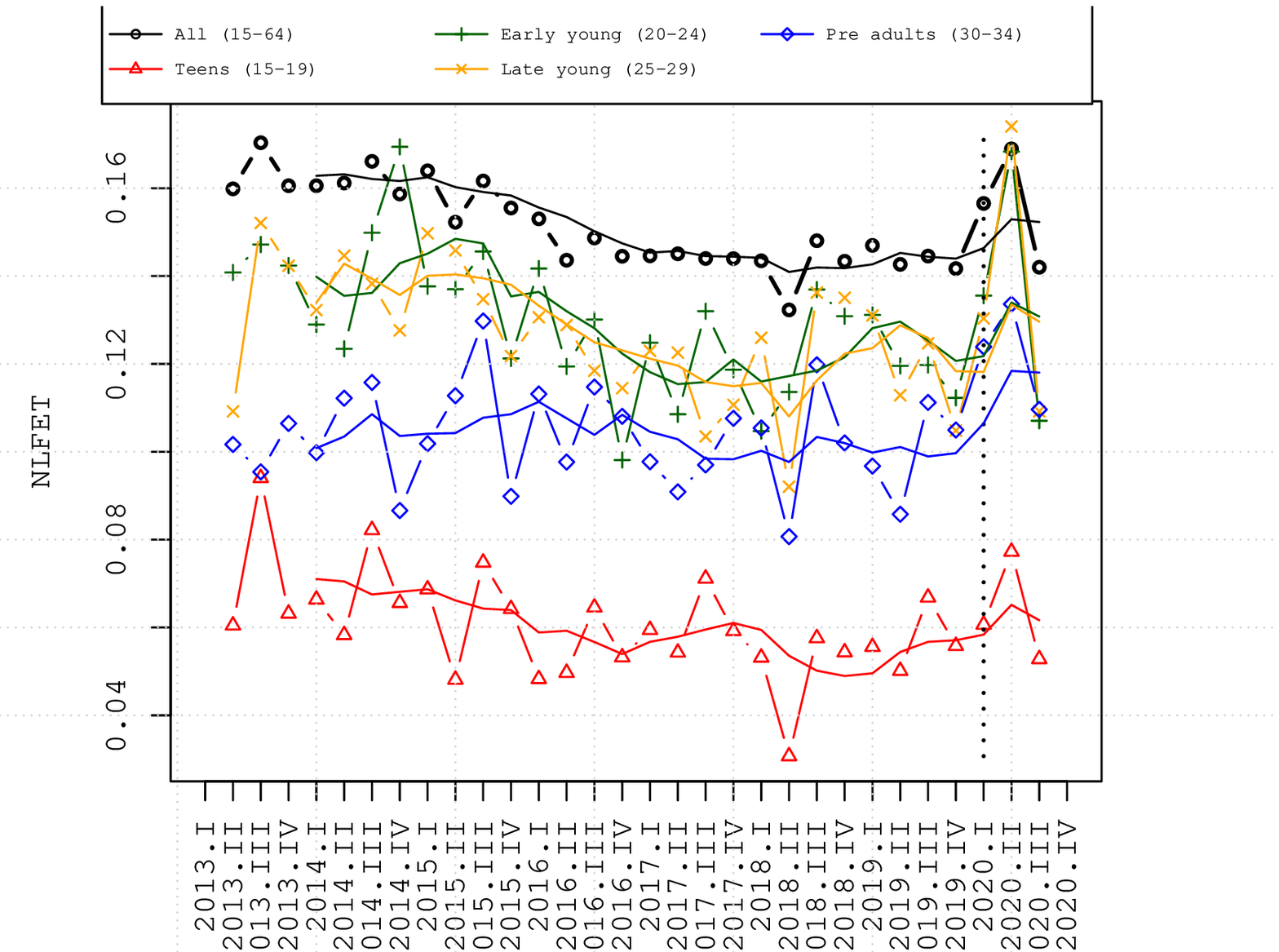}
			\caption{Males.}
				\label{fig:sharesmaleNEET}
			\vspace{0.2cm}
		\end{subfigure}
		\begin{subfigure}[t]{0.32\textwidth}
			\centering
			\includegraphics[width=1\linewidth]{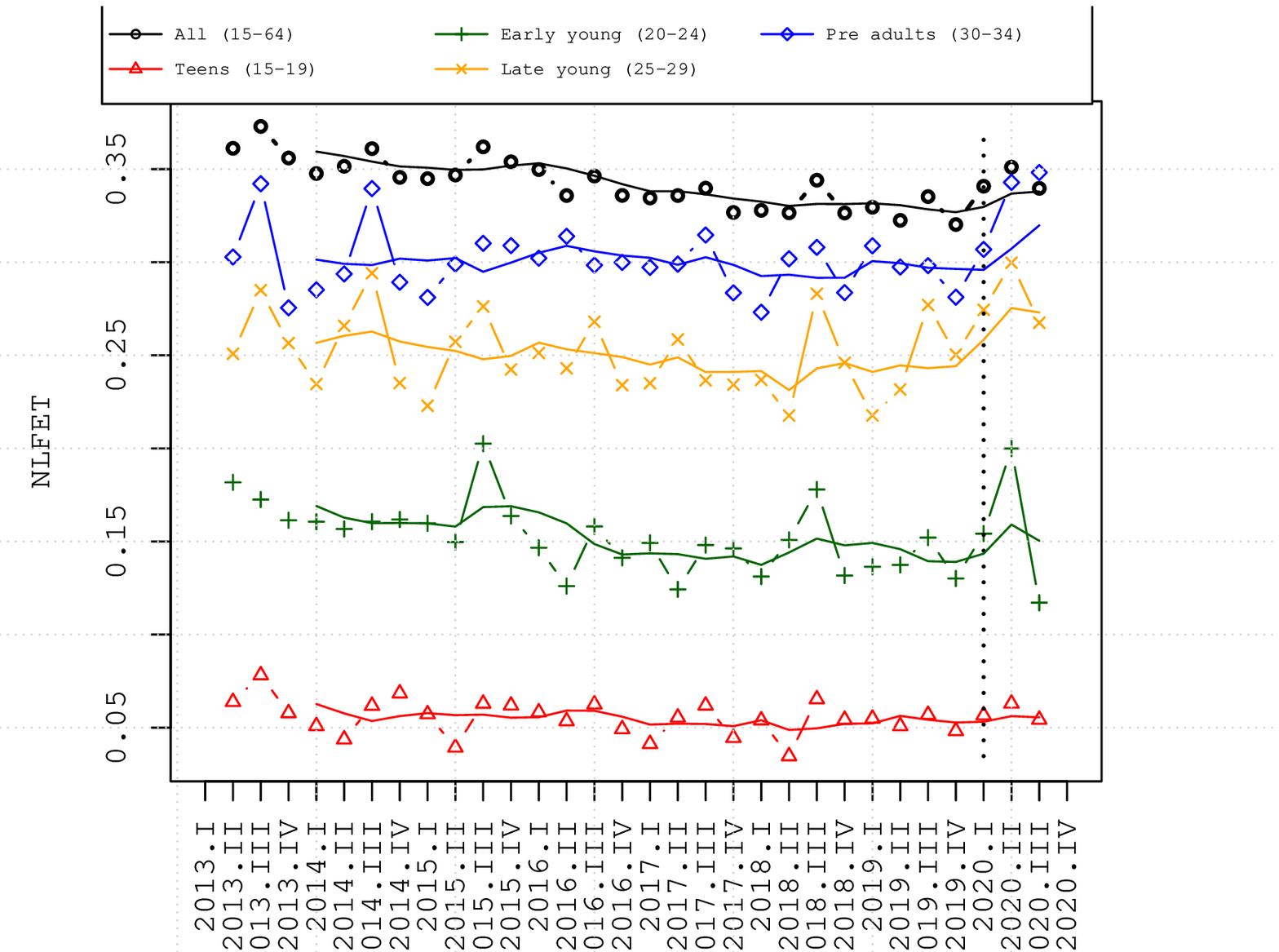}
			\caption{Females.}
				\label{fig:actualMassFemaleNLFET}
		\end{subfigure}
		\begin{subfigure}[t]{0.32\textwidth}
		\centering
		\includegraphics[width=1\linewidth]{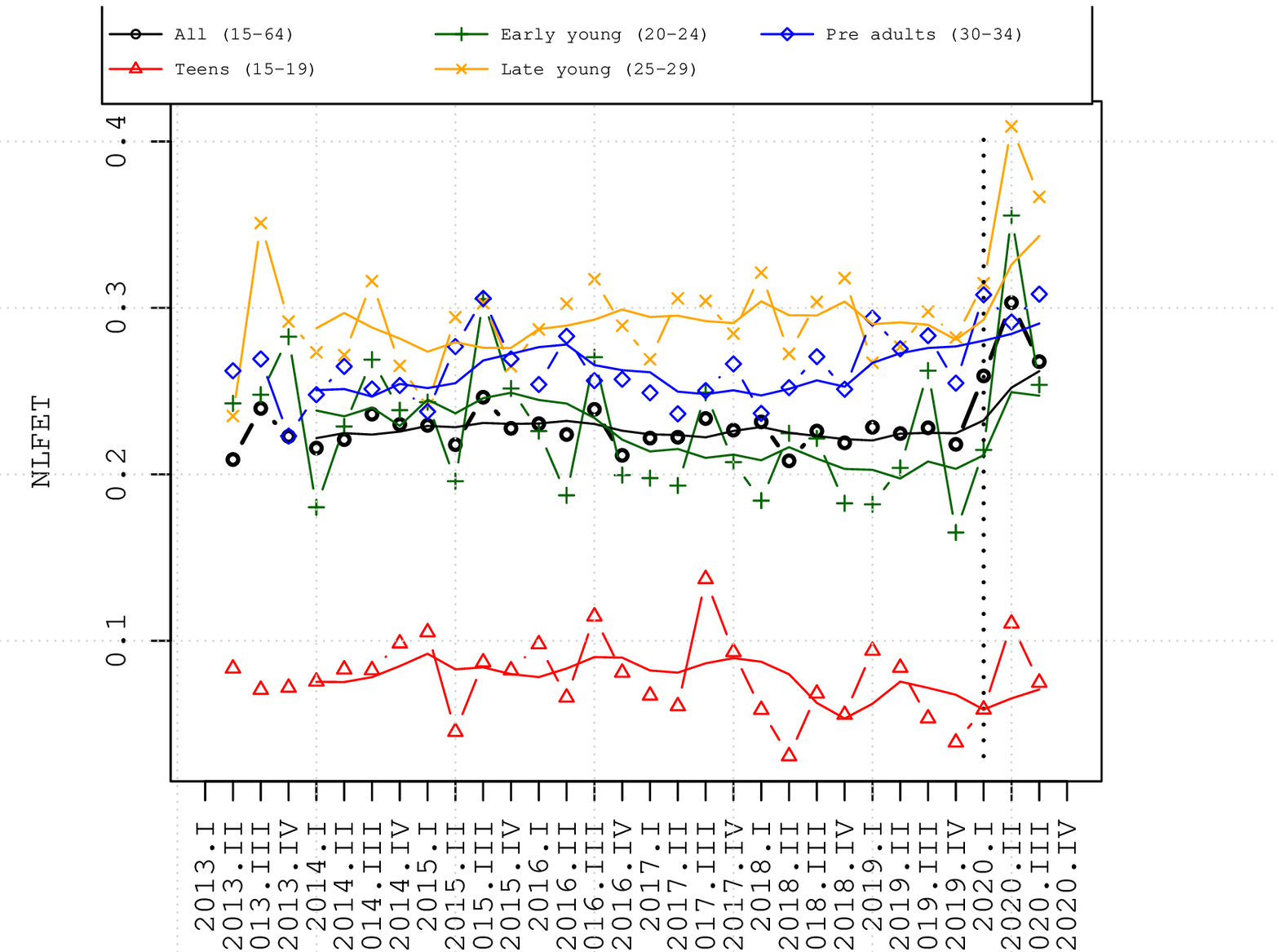}
		\caption{Non citizens.}
			\label{fig:actualMassNoCitizenshipNLFET}
	\end{subfigure}
\caption*{\scriptsize{\textit{Source}: LFS 3-month longitudinal data as provided by the Italian Institute of Statistics (ISTAT).}}
	\end{figure}

Finally, in quarter II of 2020 the share of individuals who joined the NLFET state increased across all categories. Although declining in quarter III of 2020, the share remained higher compared to the pre-pandemic value for late young, particularly females (+2\%). Higher values also persisted for female pre-adults, whose share in the NLFET state did not shrink in quarter III of 2020, marking a quite large sustained increase from 27\% pre-pandemic to 30\% (Figure \ref{fig:actualMassFemaleNLFET}). The share of non Italian citizens in the NLFET state increased and persisted at higher levels across all age categories, except for teens. Among early and late young non Italian citizens the increase in the share ranged between 3\% and 5\% (Figure \ref{fig:actualMassNoCitizenshipNLFET}).
	
\subsection{Transitions between states}\label{sec:transitionProbabilities}

In this section we compare the transition probabilities calculated in quarters II and III of 2019 (2019.II and 2019.III), before the pandemic,  and in  quarters II and III of 2020 (2020.II and 2020.III), in the midst of the pandemic, as reported in Tables \ref{tab:wholeWorkingYoung} and \ref{tab:wholeWorkingLateYoung}.

We can summarize the most important findings as follows:
(i) the transition from school to work for early young became marginally more difficult, with fewer transitions to temporary employment and more to unemployment, with an approximately stable probability of exiting the education state; on the contrary, for late young  the probability of exiting the education state increased in favor of both self-employment and unemployment;
(ii) the larger fall into the NLFET state is a shared feature of both early and late young in quarter II of 2020; however, while in quarter III of 2020 early young substantially decreased their persistence in the NLFET state and increased their transitions to education and unemployment, the late young persistence in the NLFET state only marginally decreased, largely in favor of unemployment;
and, finally, (iii) both early and late young did not change their propensity to return to school in quarter II of 2020; in quarter III of 2020 the transition from the NLFET state to education increased for early young, while the transition from unemployment to education increased for late young. Below, we document these findings with detailed references to the estimated transition probabilities reported in Tables \ref{tab:wholeWorkingYoung} and \ref{tab:wholeWorkingLateYoung}.

\paragraph{Transitions from school to work}
The first aspect to be noticed is the decrease in the persistence in the education state for late young (from 81\% in quarter III of 2019 to 76\% in quarter III of 2020), while we do not observe a significant change for the early young category. The transition rates from education to temporary and permanent employment did not change across the quarters considered for any of the two categories, being stable at around 4\% and 1\%, respectively for early young, and at 6\% and 2\%, respectively for late young. The transition rate from education to self-employment instead increased for late young from 1\% in quarter III of 2019 to 4\% in quarter III of 2020, while no change is observed among early young. For both categories the transition from education to unemployed substantially increased in quarter III of 2020, from 1\% to 4\% for  early young and from 3\% to 8\% for  late young.
 
\paragraph{Transitions into the NLFET state}
We observe an increase in the persistence in the NLFET state in quarter II of 2020 as well as a large increase in the transitions towards the NLFET state from other states for both early and late young. Specifically, the transition probability from temporary employment increased from 2\% (3\%) in quarter II of 2019 to 10\% (10\%) in quarter II of 2020 for early (late) young. Likewise, the transition probability from unemployment to the NLFET state increased from 24\% (29\%) to 52\% (48\%) for  early (late) young. Finally, we also observe that even among workers who have been furloughed, the probability to transit to the NLFET state increased from 0\% in quarter II of 2019 to 14\% in quarter II of 2020 among the late young. In quarter III of 2020 the persistence in the NLFET state decreased from 63\% to 41\% for  early young and from 67\% to 61\% for  late young. 
     
\paragraph{Transitions back to schooling}
The comparison of the transitions to education in quarter II of 2019 and 2020 shows no significant differences, except for the transition from unemployment, which significantly decreased for early young only.
On the contrary, in quarter III of 2020, we observe an increase in the flows towards education from the NLFET state for  early and  late young (from 6\% to 12\% for  early young and from 4\% to 6\% for  late young), and from unemployment for  late young alone (from 7\% to 14\%). 

\begin{table}[htbp] \centering 
 \caption{Transition probabilities of  early young in the pre-pandemic quarters II and III of 2019 (2019.II and 2019.III) and in pandemic quarters II and III of 2020 (2020.II and 2020.III).} 
  	\label{tab:wholeWorkingYoung} 
  	\begin{scriptsize}
  		\begin{tabular}{@{\extracolsep{5pt}} cccccccc} 
  			\\[-1.8ex]\hline 
  			\hline \\[-1.8ex] 
  			& SE 2019.II & TE 2019.II & PE 2019.II & U 2019.II & NLFET 2019.II & EDU 2019.II & FS 2019.II \\ 
  			\hline \\[-1.8ex] 
  		SE 2019.I & $0.78$ & $0.02$ & $0.05$ & $0$ & $0.03$ & $0.13$ & $0$ \\ 
  		TE 2019.I & $0.01$ & $0.78$ & $0.05$ & $0.06$ & ${0.02}$ & $0.08$ & $0$ \\ 
  		PE 2019.I & $0.01$ & $0.01$ & $0.92$ & $0.02$ & $0.02$ & $0.02$ & $0$ \\ 
  		U 2019.I & $0.01$ & $0.13$ & $0.02$ & $0.46$ & ${0.24}$ & $0.13$ & $0$ \\ 
  		NLFET 2019.I & $0.02$ & $0.08$ & $0.02$ & $0.21$ & ${0.59}$ & ${0.08}$ & $0$ \\ 
  		EDU 2019.I & $0.01$ & ${0.04}$ & ${0.01}$ & ${0.02}$ & $0.03$ & $0.88$ & $0$ \\ 
  		FS 2019.I & $0.14$ & $0.14$ & $0.14$ & $0.14$ & $0.14$ & $0.14$ & $0.14$ \\ 
  			\hline \\[-1.8ex] 
  			
  			& SE 2019.III & TE 2019.III & PE 2019.III & U 2019.III & NLFET 2019.III & EDU 2019.III & FS 2019.III \\ 
  			\hline \\[-1.8ex] 
  	SE 2019.II & $0.72$ & $0.04$ & $0.03$ & $0.02$ & $0.04$ & $0.15$ & $0$ \\ 
  	TE 2019.II & $0$ & $0.79$ & $0.07$ & $0.03$ & ${0.05}$ & $0.06$ & $0$ \\ 
  	PE 2019.II & $0$ & $0.04$ & $0.89$ & $0.01$ & $0.01$ & $0.04$ & $0$ \\ 
  	U 2019.II & $0.02$ & $0.17$ & $0.01$ & $0.36$ & ${0.31}$ & $0.13$ & $0$ \\ 
  	NLFET 2019.II & $0.02$ & $0.10$ & $0.04$ & $0.16$ & ${0.63}$ & ${0.06}$ & $0$ \\ 
  	EDU 2019.II & $0.01$ & ${0.06}$ & ${0.01}$ & ${0.01}$ & $0.03$ & $0.88$ & $0$ \\ 
  	FS 2019.II & $0.14$ & $0.14$ & $0.14$ & $0.14$ & $0.14$ & $0.14$ & $0.14$ \\
  			\hline \\[-1.8ex]  
  			& SE 2020.II & TE 2020.II & PE 2020.II & U 2020.II & NLFET 2020.II & EDU 2020.II & FS 2020.II \\ 
  			\hline \\[-1.8ex] 
  		SE 2020.I & $0.64$ & $0.01$ & $0.10$ & $0.05$ & $0.05$ & $0.15$ & $0$ \\ 
  		TE 2020.I & $0.01$ & $0.66$ & $0.06$ & $0.07$ & ${0.10}$ & $0.06$ & $0.05$ \\ 
  		PE 2020.I & $0.03$ & $0.06$ & $0.65$ & $0.03$ & $0.04$ & $0.02$ & $0.18$ \\ 
  		U 2020.I & $0.01$ & $0.07$ & $0.02$ & $0.29$ & ${0.52}$ & ${0.07}$ & $0.01$ \\ 
  		NLFET 2020.I & $0.01$ & $0.04$ & $0.01$ & $0.19$ & ${0.69}$ & $0.05$ & $0$ \\ 
  		EDU 2020.I & $0.01$ & ${0.03}$ & ${0.01}$ & ${0.01}$ & $0.03$ & $0.91$ & $0.01$ \\ 
  		FS 2020.I & $0$ & $0$ & $1$ & $0$ & $0$ & $0$ & $0$ \\
  			\hline \\[-1.8ex] 
  			
  			& SE 2020.III & TE 2020.III & PE 2020.III & U 2020.III & NLFET 2020.III & EDU 2020.III & FS 2020.III \\ 
  			\hline \\[-1.8ex] 
  		SE 2020.II & $0.75$ & $0.09$ & $0.01$ & $0.02$ & $0.02$ & $0.11$ & $0$ \\ 
  		TE 2020.II & $0.01$ & $0.80$ & $0.06$ & $0.04$ & ${0.03}$ & $0.07$ & $0$ \\ 
  		PE 2020.II & $0.01$ & $0.05$ & $0.85$ & $0.02$ & $0.02$ & $0.03$ & $0.02$ \\ 
  		U 2020.II & $0.02$ & $0.12$ & $0.02$ & $0.41$ & ${0.32}$ & $0.10$ & $0$ \\ 
  		NLFET 2020.II & $0$ & $0.13$ & $0.01$ & $0.32$ & ${0.41}$ & ${0.12}$ & $0$ \\ 
  		EDU 2020.II & $0.01$ & ${0.04}$ & ${0.01}$ & ${0.04}$ & $0.02$ & $0.87$ & $0$ \\ 
  		FS 2020.II & $0$ & $0.04$ & $0.75$ & $0.05$ & $0.06$ & $0.02$ & $0.08$ \\  
  			\hline \\[-1.8ex] 
  		\end{tabular} 
  	\end{scriptsize}
  	\caption*{\scriptsize{\textit{Note}:  The transition probabilities are computed as  the ratio of the number of young (age 20-24) who were in state $i$ at time $t$ and in state $j$ at time $t+1$ and the number of individuals in state $i$ at time $t$, where $i,j \in \{1,2,..7\}$ . SE refers to self-employment, TE to temporary employment, PE to permanent employment, U to unemployment, NLFET to not in the labour force, employment or training, EDU education and FS to furlough scheme. Source: LFS 3-month longitudinal data as provided by the Italian Institute of Statistics (ISTAT).}}
  \end{table} 
 
\begin{table}[htbp] \centering 
\caption{Transition probabilities of late young in the pre-pandemic quarters II and III of 2019 (2019.II and 2019.III) and in the pandemic quarters II and III of 2020 (2020.II and 2020.III)} 
\label{tab:wholeWorkingLateYoung} 
\begin{scriptsize}
\begin{tabular}{@{\extracolsep{5pt}} cccccccc} 
\\[-1.8ex]\hline 
\hline \\[-1.8ex] 
& SE 2019.II & TE 2019.II & PE 2019.II & U 2019.II & NLFET2019.II & EDU 2019.II & FS 2019.II \\ 
\hline \\[-1.8ex] 
SE 2019.I & $0.88$ & $0.04$ & $0.02$ & $0.02$ & $0.04$ & ${0.02}$ & $0$ \\ 
TE 2019.II & $0.01$ & $0.83$ & $0.07$ & $0.03$ & ${0.03}$ & ${0.02}$ & $0$ \\ 
PE 2019.II & $0.01$ & $0.03$ & $0.93$ & $0.01$ & $0.02$ & $0.01$ & $0$ \\ 
U 2019.II & $0.03$ & $0.11$ & $0.04$ & $0.46$ & ${0.29}$ & ${0.07}$ & $0$ \\ 
NLFET2019.II & $0.01$ & $0.07$ & $0.03$ & $0.21$ & ${0.63}$ & $0.06$ & $0$ \\ 
EDU 2019.II & ${0.01}$ & ${0.06}$ & ${0.02}$ & ${0.05}$ & $0.04$ & $0.82$ & $0$ \\ 
FS 2019.II & $0$ & $0$ & $0.36$ & $0$ & ${0}$ & $0$ & $0.64$ \\ 
\hline \\[-1.8ex] 

 & SE 2019.III & TE 2019.III & PE 2019.III & U 2019.III & NLFET2019.III & EDU 2019.III & FS 2019.III \\ 
\hline \\[-1.8ex] 
SE 2019.II & $0.87$ & $0.04$ & $0.01$ & $0.04$ & $0.04$ & ${0.01}$ & $0$ \\ 
TE 2019.II & $0.01$ & $0.77$ & $0.09$ & $0.04$ & ${0.07}$ & ${0.02}$ & $0$ \\ 
PE 2019.II & $0.01$ & $0.03$ & $0.94$ & $0$ & $0.02$ & $0$ & $0$ \\ 
U 2019.II & $0.02$ & $0.12$ & $0.03$ & $0.40$ & ${0.37}$ & ${0.07}$ & $0$ \\ 
NLFET2019.II & $0.02$ & $0.08$ & $0.01$ & $0.19$ & ${0.67}$ & $0.04$ & $0$ \\ 
EDU 2019.II & ${0.01}$ & ${0.06}$ & ${0.01}$ & ${0.03}$ & $0.07$ & $0.81$ & $0$ \\ 
FS 2019.II & $0.14$ & $0.14$ & $0.14$ & $0.14$ & ${0.14}$ & $0.14$ & $0.14$ \\ 
\hline \\[-1.8ex]  
 & SE 2020.II & TE 2020.II & PE 2020.II & U 2020.II & NLFET2020.II & EDU 2020.II & FS 2020.II \\ 
\hline \\[-1.8ex] 
SE 2020.I & $0.88$ & $0.02$ & $0.03$ & $0.02$ & $0.01$ & ${0.02}$ & $0.01$ \\ 
TE 2020.I & $0.02$ & $0.68$ & $0.06$ & $0.07$ & ${0.10}$ & ${0.04}$ & $0.04$ \\ 
PE 2020.I & $0$ & $0.04$ & $0.76$ & $0$ & $0.03$ & $0.01$ & $0.16$ \\ 
U 2020.I & $0.01$ & $0.09$ & $0.03$ & $0.31$ & ${0.48}$ & ${0.06}$ & $0.02$ \\ 
NLFET2020.I & $0.01$ & $0.03$ & $0.02$ & $0.16$ & ${0.74}$ & $0.03$ & $0$ \\ 
EDU 2020.I & ${0.03}$ & ${0.05}$ & ${0.02}$ & ${0.05}$ & $0.06$ & $0.79$ & $0$ \\ 
FS 2020.I & $0$ & $0.01$ & $0.84$ & $0$ & ${0.15}$ & $0$ & $0$ \\ 
\hline \\[-1.8ex] 
		
& SE 2020.III & TE 2020.III & PE 2020.III & U 2020.III & NLFET 2020.III & EDU 2020.III & FS 2020.III \\ 
\hline \\[-1.8ex] 
SE 2020.II & $0.86$ & $0.02$ & $0.03$ & $0.03$ & $0.02$ & ${0.04}$ & $0$ \\ 
TE 2020.II & $0.02$ & $0.76$ & $0.06$ & $0.05$ & ${0.09}$ & ${0.02}$ & $0$ \\ 
PE 2020.II & $0.02$ & $0.02$ & $0.94$ & $0.01$ & $0.01$ & $0$ & $0.01$ \\ 
U 2020.II & $0.03$ & $0.09$ & $0.01$ & $0.46$ & ${0.28}$ & ${0.14}$ & $0$ \\ 
NLFET 2020.II & $0$ & $0.07$ & $0.02$ & $0.23$ & ${0.61}$ & $0.06$ & $0$ \\ 
EDU 2020.II & ${0.04}$ & ${0.06}$ & ${0.01}$ & ${0.08}$ & $0.06$ & $0.76$ & $0$ \\ 
FS 2020.II & $0$ & $0.10$ & $0.76$ & $0.01$ & ${0.05}$ & $0$ & $0.07$ \\ 
\hline \\[-1.8ex] 
\end{tabular} 
\end{scriptsize}
\caption*{\scriptsize{\textit{Note}:  The transition probabilities are computed as the ratio of the number of late young (age 25-29) who were in state $i$ at time $t$ and in state $j$ at time $t+1$ and the number of individuals in state $i$ at time $t$, where $i,j \in \{1,2,..7\}$ . SE refers to self-employment, TE to temporary employment, PE to permanent employment, U to unemployment, NLFET to not in education, employment or training, EDU education and FS to furlough scheme. Source: LFS 3-month longitudinal data as provided by the Italian Institute of Statistics (ISTAT). }}
\end{table} 

\subsection{Duration of the transition from school to work}\label{sec:FPT}

\begin{figure}[htbp]
	\caption{First passage time (FPT) to permanent employment for  early young by different categories (all, males, females, non Italian citizens and living in the South).}
	\label{fig:firstTimePassagePEYoung}
	\vspace{0.3cm}
	\begin{subfigure}[b]{0.32\textwidth}
		\centering
		\includegraphics[width=1\linewidth]{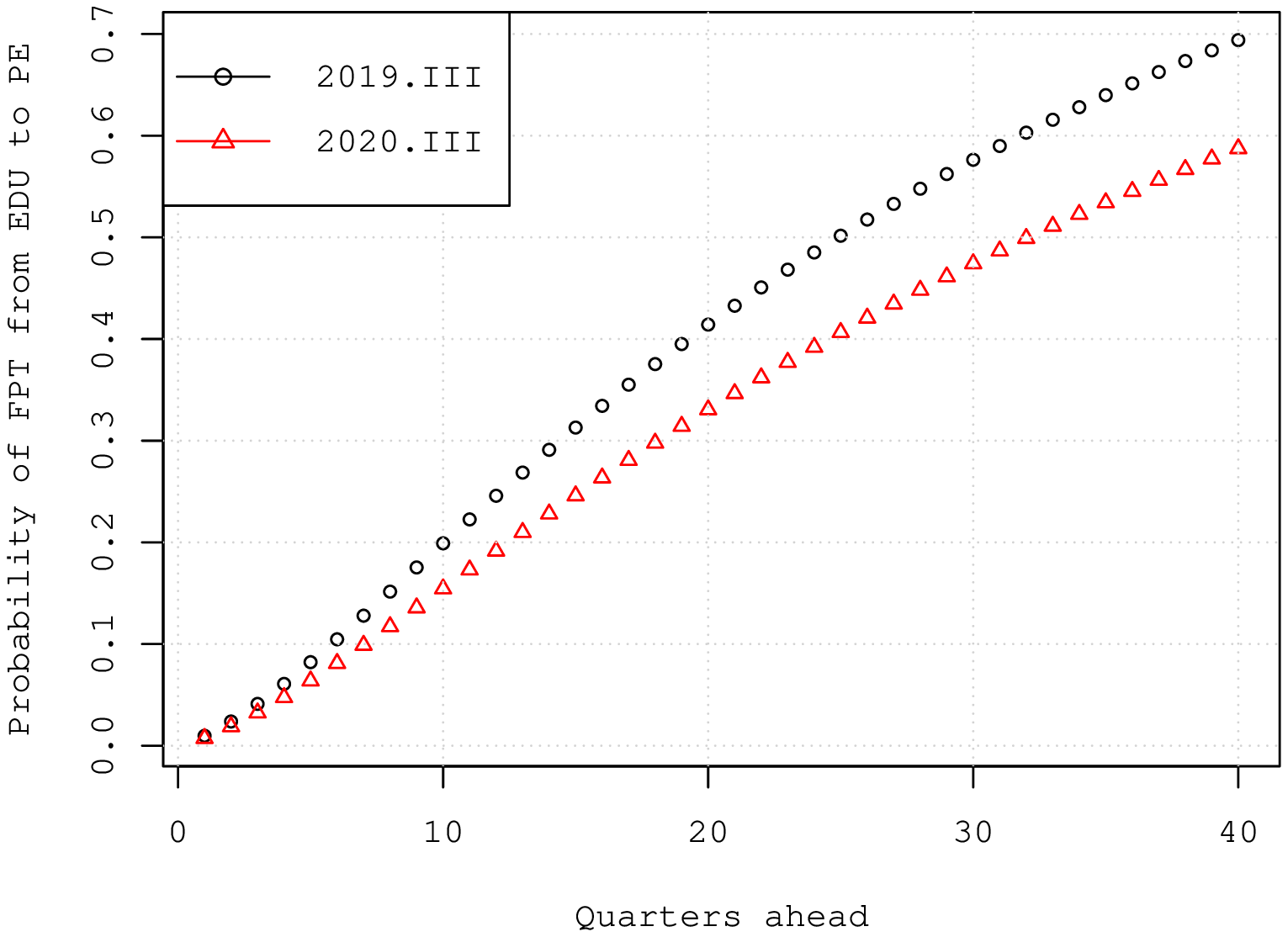}
		\caption{All.}
		\label{fig:firstTimePassageFromEDUtoPEAllYoung}
		\vspace{0.2cm}
	\end{subfigure}
	\begin{subfigure}[b]{0.32\textwidth}
		\centering
		\includegraphics[width=1\linewidth]{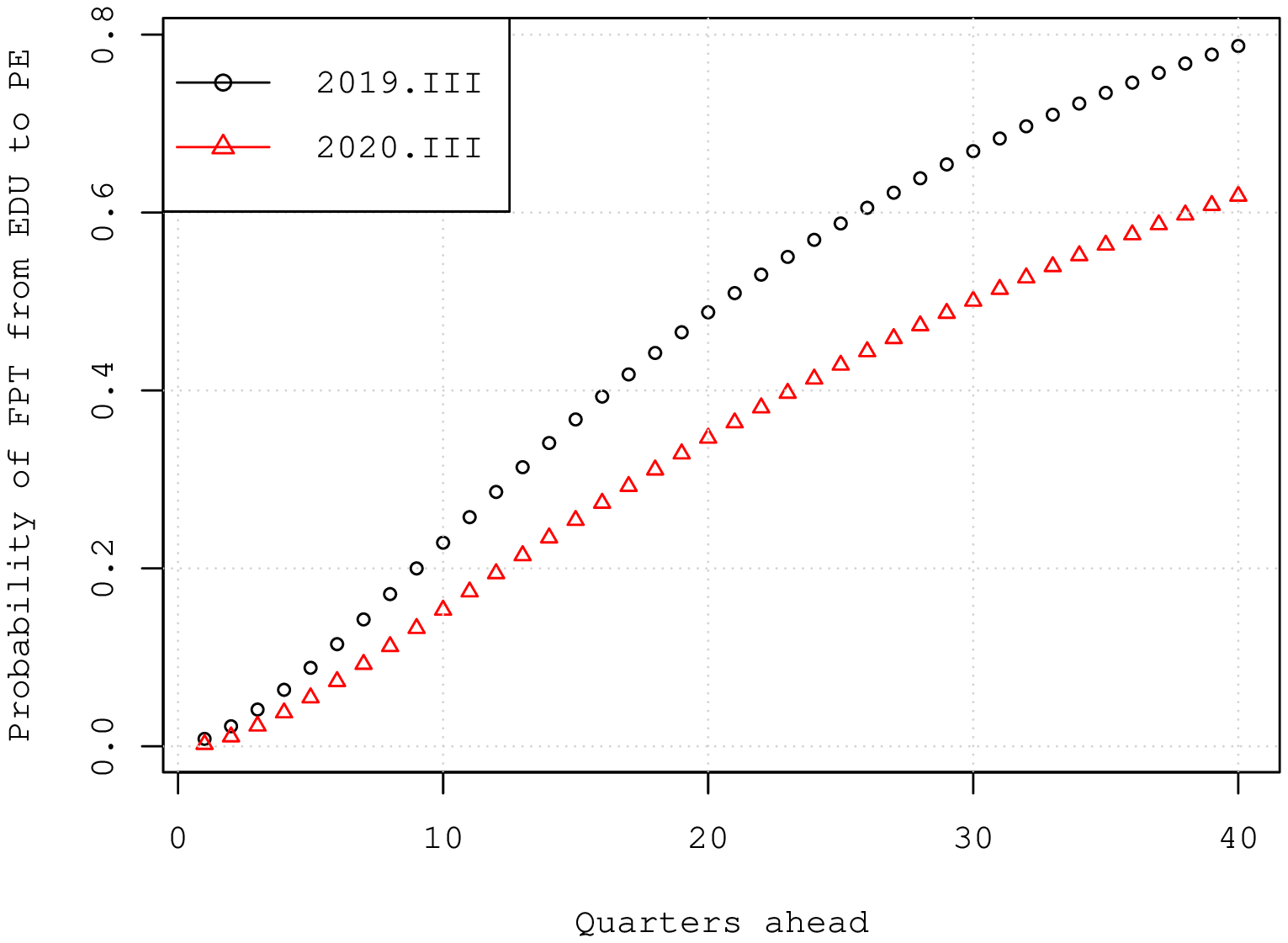}
		\caption{Males.}
		\label{fig:firstTimePassageFromEDUtoPEMaleYoung}
	\end{subfigure}
	\begin{subfigure}[b]{0.32\textwidth}
		\centering
		\includegraphics[width=1\linewidth]{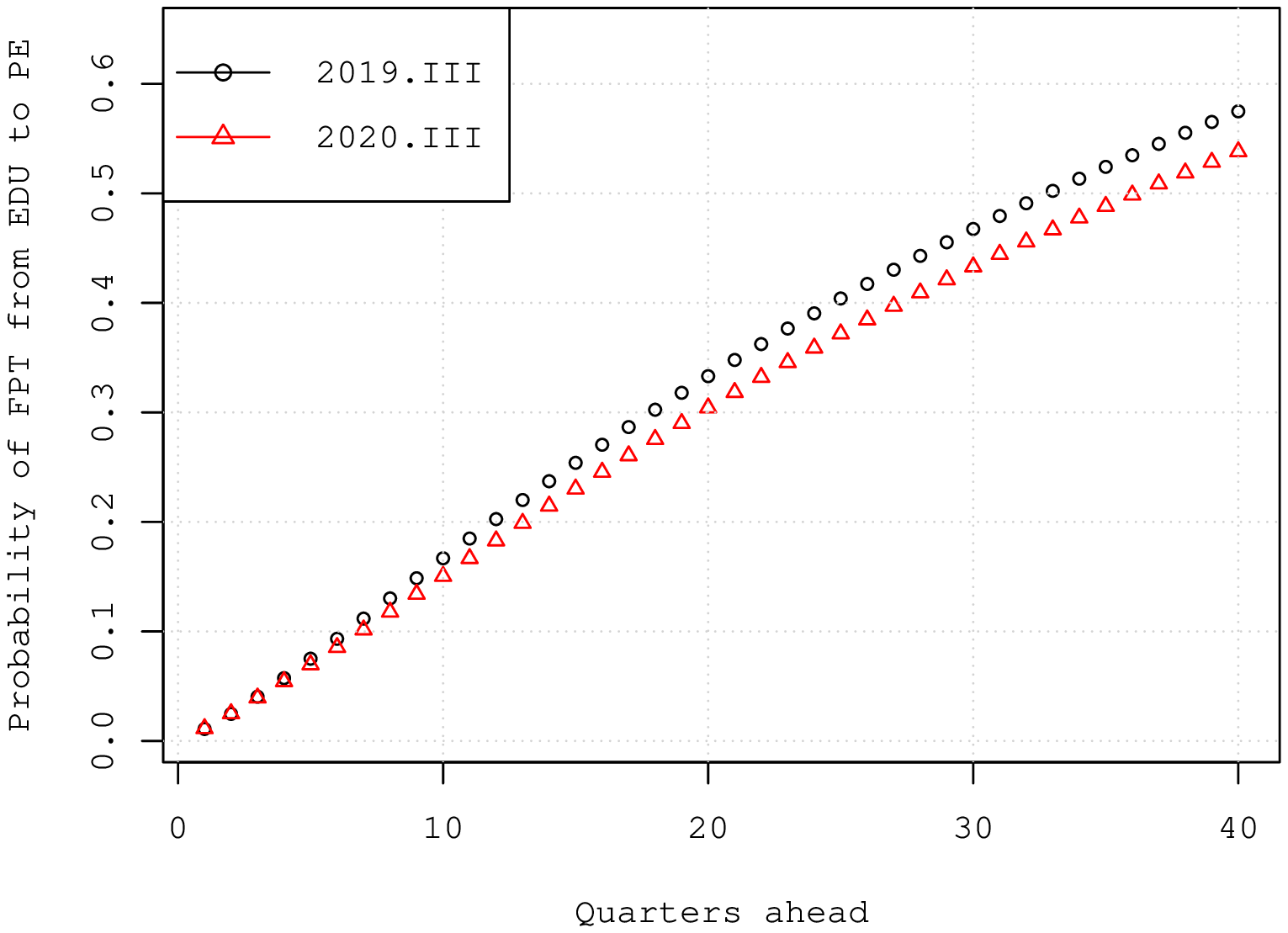}
		\caption{Females.}
		\label{fig:firstTimePassageFromEDUtoPEFemaleYoung}
		\vspace{0.2cm}
	\end{subfigure}
	\begin{subfigure}[b]{0.32\textwidth}
		\centering
		\includegraphics[width=1\linewidth]{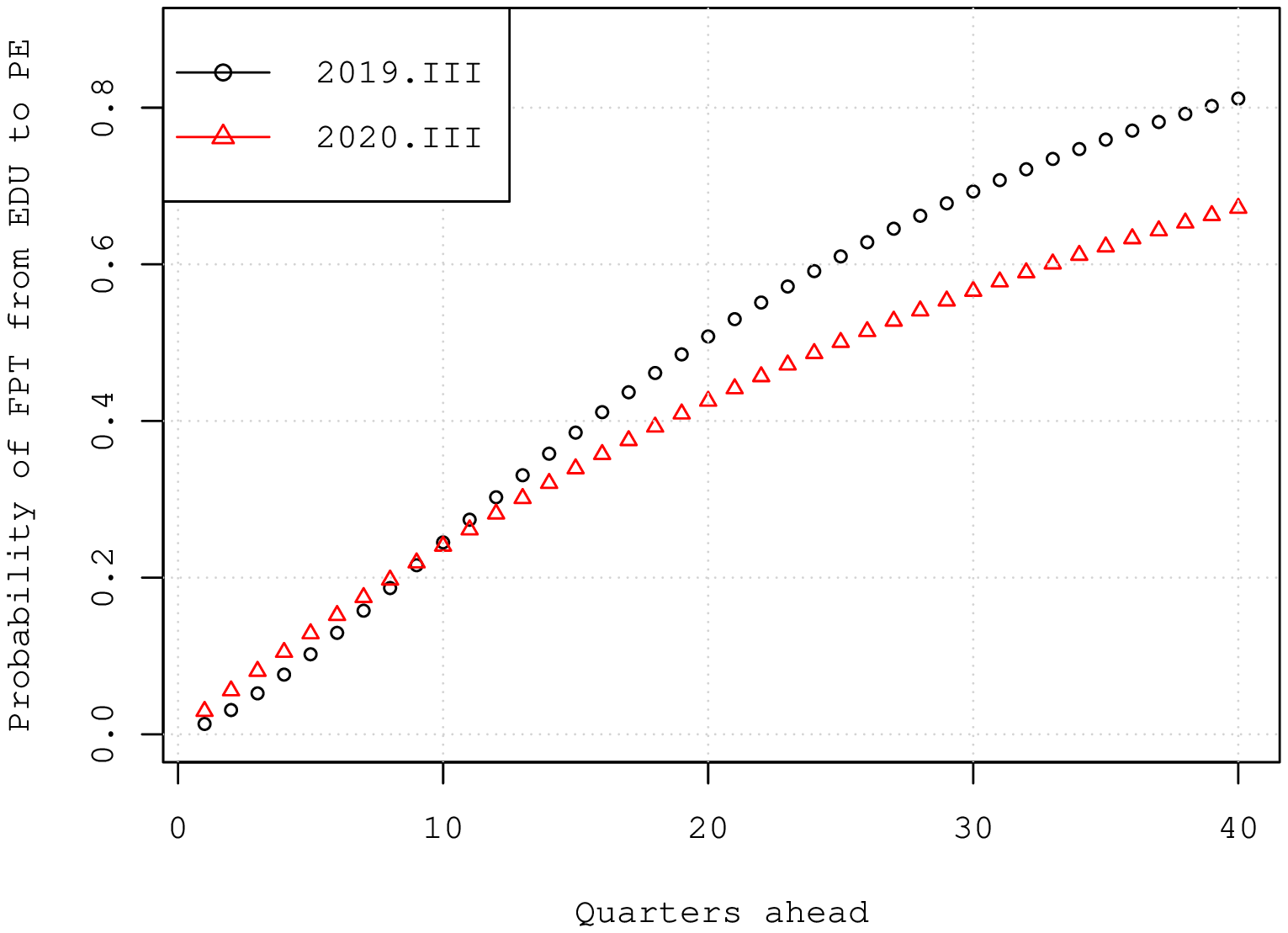}
		\caption{Non Italian citizens.}
		\label{fig:firstTimePassageFromEDUtoPENoCitizYoung}
	\end{subfigure}
	\begin{subfigure}[b]{0.32\textwidth}
		\centering
		\includegraphics[width=1\linewidth]{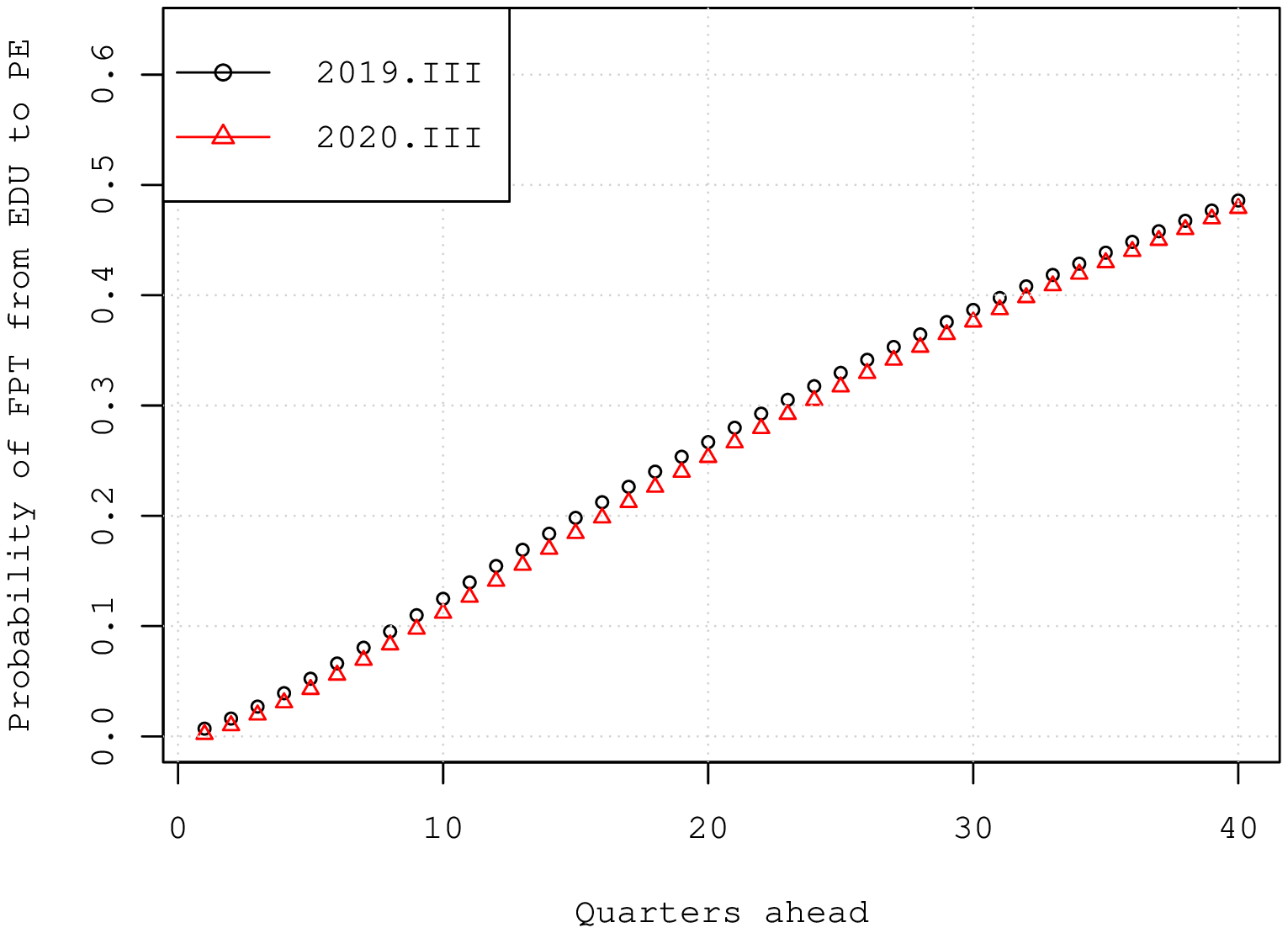}
		\caption{South.}
		\label{fig:firstTimePassageFromEDUtoPESouthYoung}
	\end{subfigure}
\caption*{\scriptsize{Source: LFS 3-month longitudinal data as provided by the Italian Institute of Statistics (ISTAT).}}
\end{figure}

In this section we discuss the impact of the pandemic on the duration of the transition from school to work (STWT) which, in our approach based on transition probabilities, it is proxied by the first passage time (FPT), i.e., the distribution of quarters which an individual takes to arrive for the first time at state $j$ starting from state $i$, and, more precisely, by the expected first passage time (EFPT), i.e., the expected number of quarters (years) it takes to go from state $i$ to state $j$ for the first time \citep[p. 818]{lieberman2001introduction} (see Appendix \ref{app:firstpassagetime} for technical details). In particular, we are interested in computing the FPT and the EFPT from education to employment, both temporary and permanent.
Figure \ref{fig:firstTimePassagePEYoung} reports the FPT from education to permanent employment for early young divided into different categories (all, males, females, non Italian citizens and living in the South). In particular, the FPT in quarter III of 2019 is calculated taking the estimated transition probabilities for that quarter and applying the procedure described in Appendix \ref{app:firstpassagetime}. The same procedure is used  to compute the FPT in quarter III of 2020. Table \ref{tab:timeuntilpassage} reports the corresponding EFPT to employment (permanent and temporary) of early young individuals in  education.

\begin{table}[htbp] \centering 
	\caption{Expected first passage time (EFPT) in years to employment (permanent and temporary) of early young individuals starting from the education state.} 
	\label{tab:timeuntilpassage} 
	\begin{scriptsize}
		\begin{tabular}{@{\extracolsep{5pt}} lcccc} 
			\\[-1.8ex]\hline 
			\hline \\[-1.8ex] 
					&\multicolumn{2}{c}{Permanent Employment}  & \multicolumn{2}{c}{Temporary Employment}  \\ 
					\hline\\[-1.8ex] 
						& 2019.III & 2020.III  & 2019.III & 2020.III \\ 
					\hline\\[-1.8ex] 
		All	& 8.63 & 11.25&3.72&4.16\\
		Males & 6.88 & 10.39&3.46&3.60\\
		Females &11.70&12.85&4.08&4.89\\
		Non Italian citizens&6.43&9.09&6.40&7.55\\
		South &14.92& 14.88 &4.23&5.98\\ 
			\hline \\[-1.8ex] 
		\end{tabular} 
	\end{scriptsize}
	\caption*{\scriptsize{\textit{Note}:  The category `non Italian citizens' is defined according to the citizenship and not country of birth as a large number of young people in Italy are second generation immigrants born in Italy. Source: LFS 3-month longitudinal data as provided by the Italian Institute of Statistics (ISTAT).}}
\end{table} 

Before the pandemic females and early young living in the South were the categories for which the STWT duration was longer \citep{pastore2014youth} (Figure \ref{fig:firstTimePassagePEYoung}). In particular, early young in education in quarter III of 2019 had 70\% probability to have been at least once in permanent employment  after 40 quarters (10 years). However, when looking at specific categories of early young, this probability ranges from 80\% for males, less than 60\% for females and only 50\% for individuals living in the South, while no difference emerges between Italian and non Italian citizens. These figures reflect in different EFPTs, which vary from 6.88 years for  males, to 11.7 for  females and 14.92 for residents in the South (Table \ref{tab:timeuntilpassage}).
The pandemic significantly affected the FPT: after 40 quarters the same probability in  quarter III of 2020 fell to 60\% from 70\%, with the corresponding EFPT spiking up to 11.25 years from 8.63 (Table \ref{tab:timeuntilpassage}). The effect of the pandemic however has have highly asymmetric, affecting mostly  males and  non Italian citizens: the EFPT raised from 6.88 years to 10.39 years for males and from 6.43 years to 9.09 years for  non Italian citizens.

\begin{figure}[htbp]
	\caption{First passage time (FPT) of early young individuals to temporary employment.}
	\label{fig:firstTimePassageFTYoung}
	\vspace{0.3cm}
	\begin{subfigure}[b]{0.32\textwidth}
		\centering
		\includegraphics[width=1\linewidth]{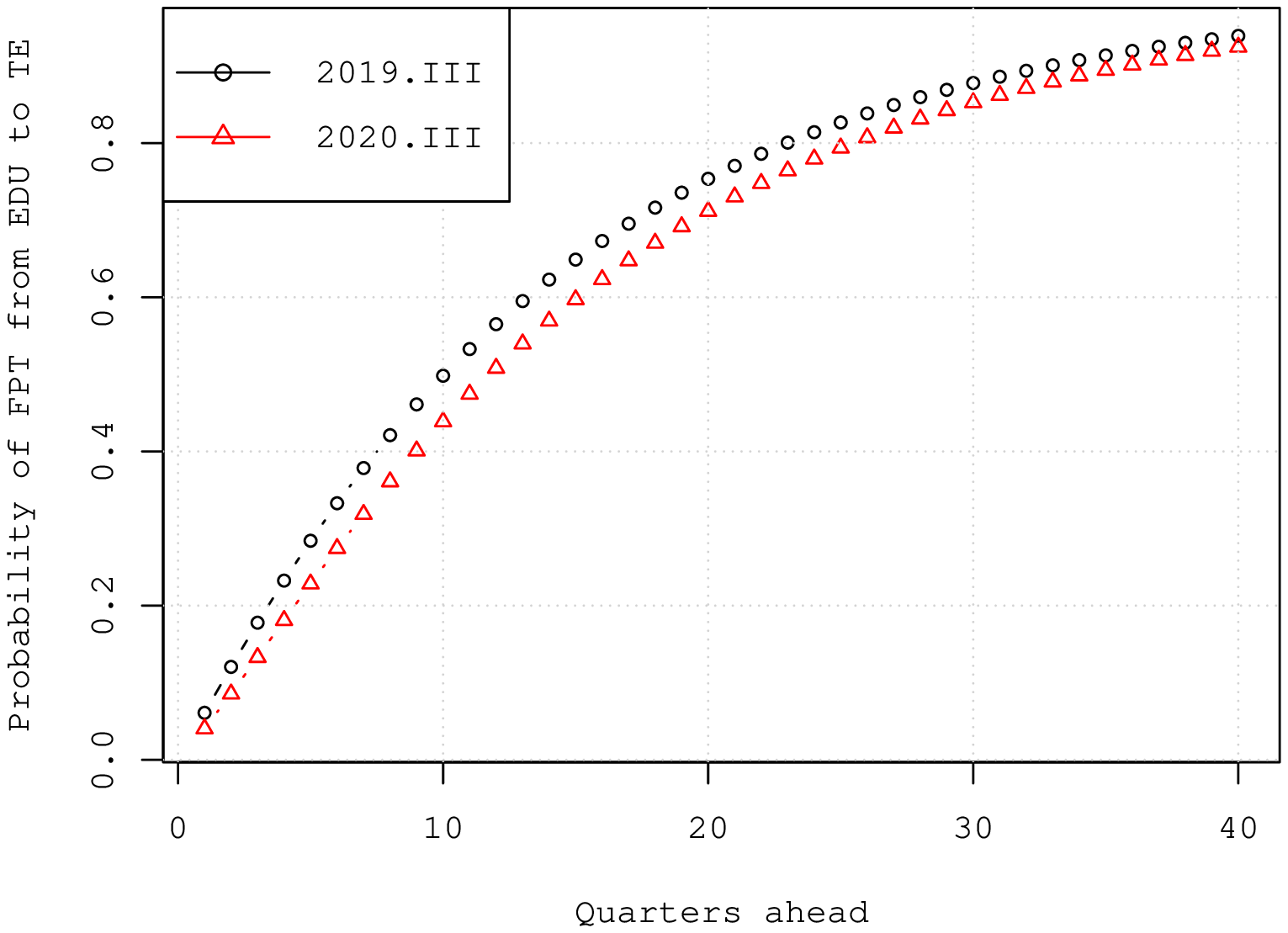}
		\caption{All.}
		\label{fig:EDU}
		\vspace{0.2cm}
	\end{subfigure}
	\begin{subfigure}[b]{0.32\textwidth}
		\centering
		\includegraphics[width=1\linewidth]{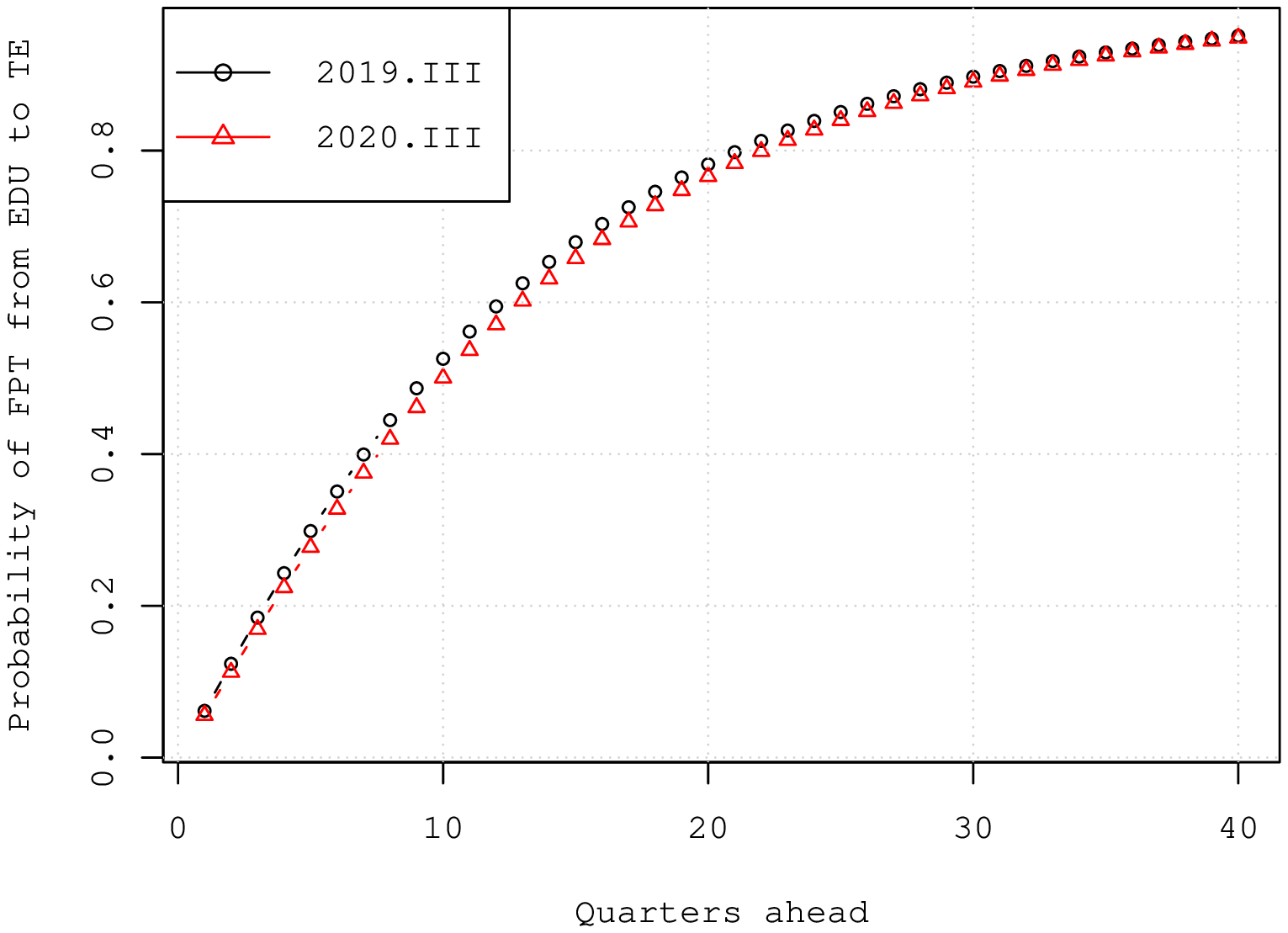}
		\caption{Males.}
		\label{fig:PE}
	\end{subfigure}
	\begin{subfigure}[b]{0.32\textwidth}
		\centering
		\includegraphics[width=1\linewidth]{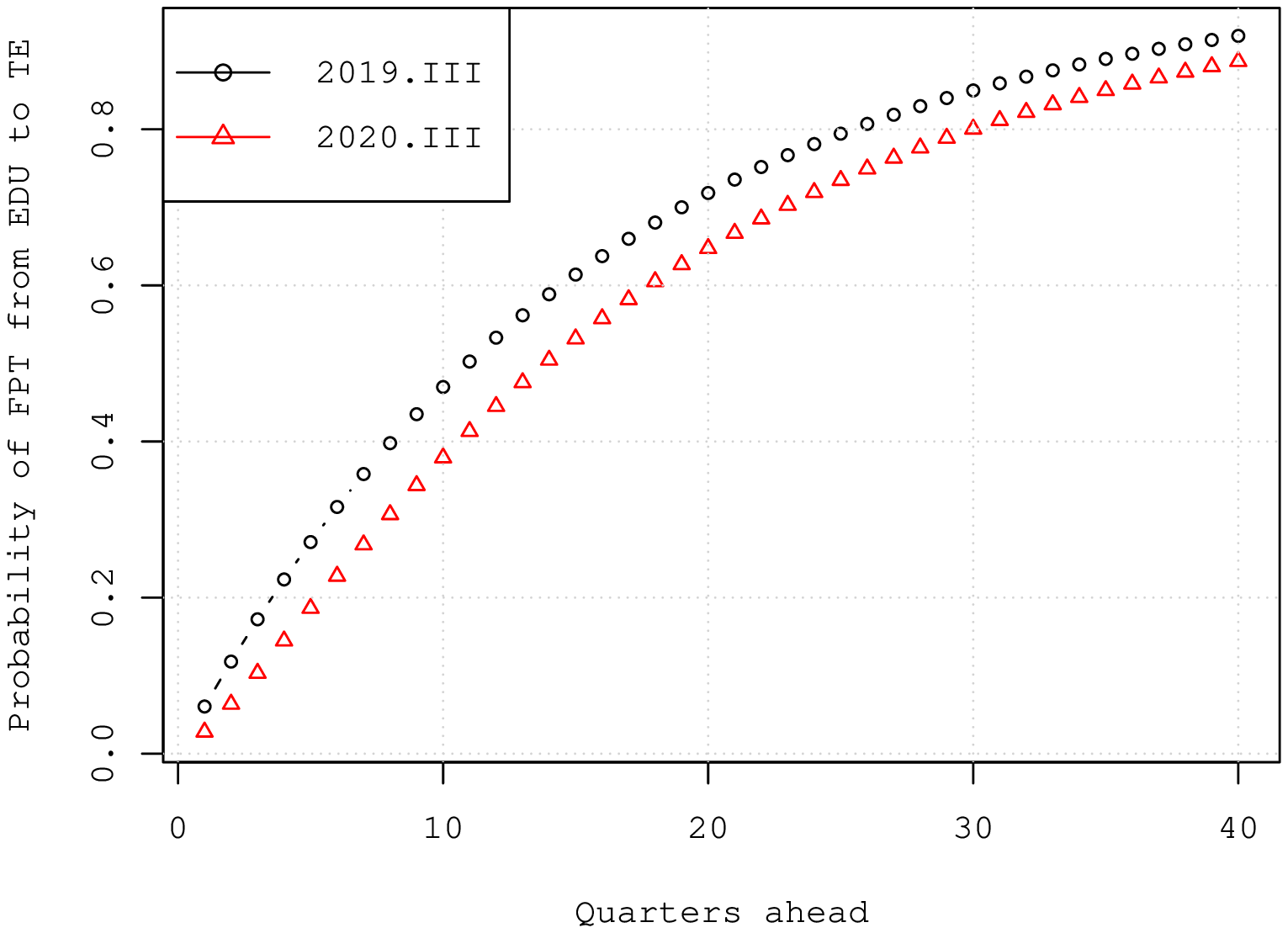}
		\caption{Females.}
		\label{fig:firstTimePassageFromEDUtoFTFemaleYoung}
		\vspace{0.2cm}
	\end{subfigure}
	\begin{subfigure}[b]{0.32\textwidth}
		\centering
		\includegraphics[width=1\linewidth]{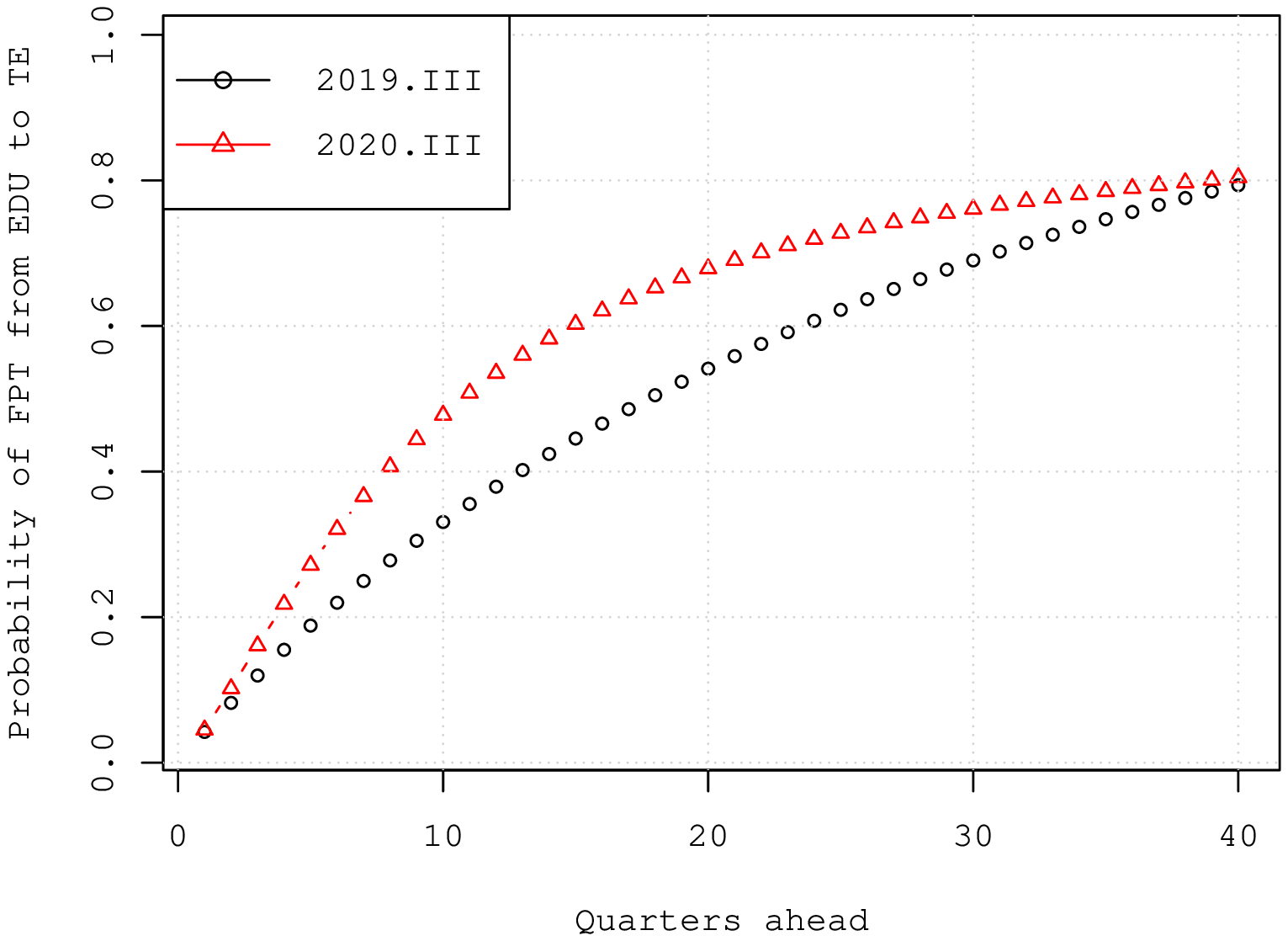}
		\caption{Non citizens.}
		\label{fig:U}
	\end{subfigure}
	\begin{subfigure}[b]{0.32\textwidth}
		\centering
		\includegraphics[width=1\linewidth]{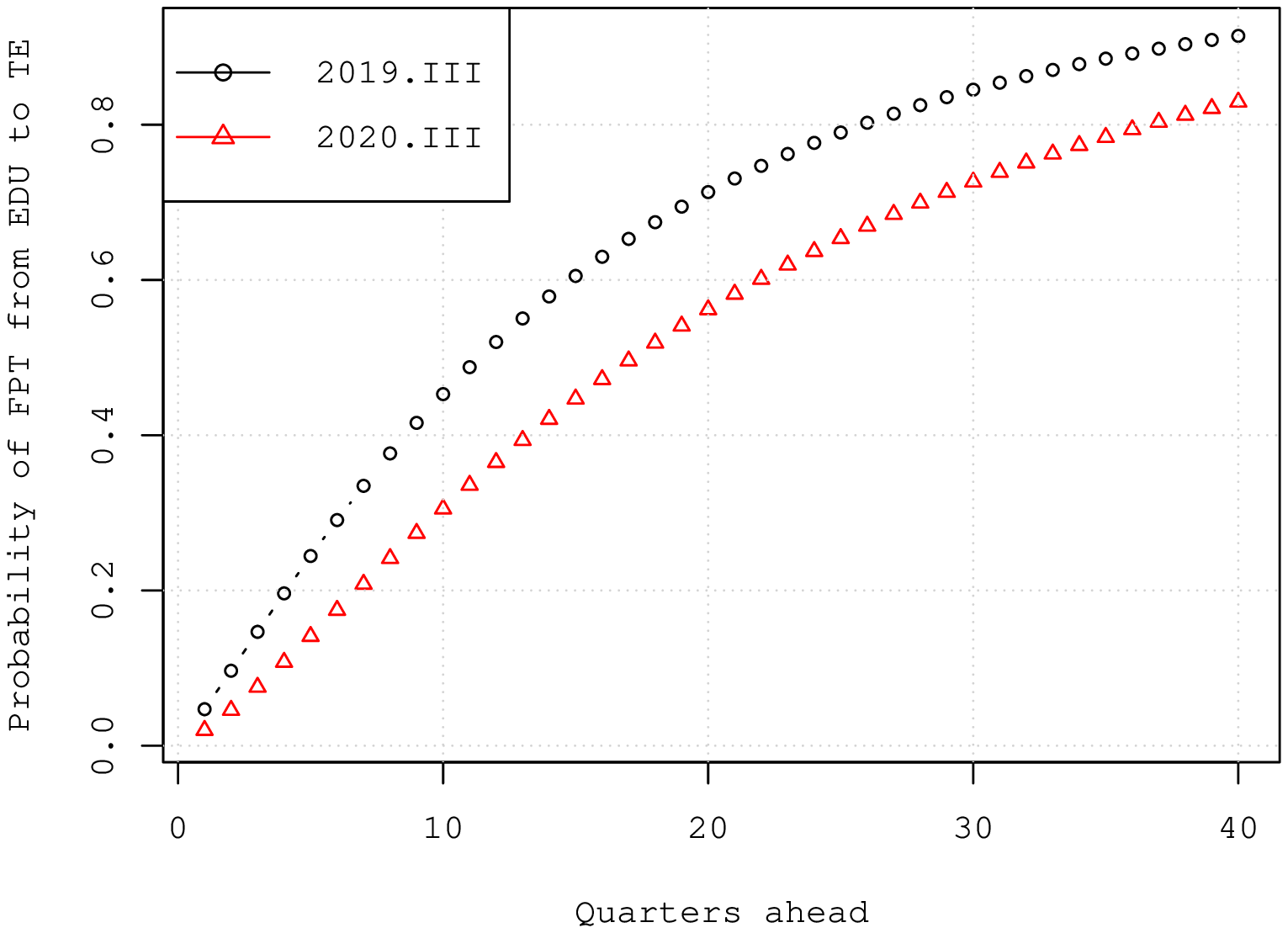}
		\caption{South.}
		\label{fig:NEET}
	\end{subfigure}
\caption*{\scriptsize{Source: LFS 3-month longitudinal data as provided by the Italian Institute of Statistics (ISTAT).}}
\end{figure}

The FPT of early young to temporary employment is similar across all categories of individuals considered (males, females, living in the South), except for non Italian citizens, for whom it is slightly lower (Figure \ref{fig:firstTimePassageFTYoung}). After 40 quarters 90\% of early young in  education in quarter II of 2019 were expected to have had at least one experience of temporary employment. This corresponds to an EFPT of approximately 4 years for all categories, with the exception of non Italian citizens, whose EFPT was approximately 6.4 years (Table \ref{tab:timeuntilpassage}).
The negative impact of the pandemic on the FPT to temporary employment has been much smaller compared to the one to permanent employment: it increased from 3.72 years in quarter III of 2019 to 4.16 years in quarter III of 2020. Residents in the South have been  affected the most with an EFPT to temporary employment increasing from 4.23 years to 5.98 years.

\section{Concluding remarks \label{sec:concludingRemarks}}
	
In this paper we provide evidence of the extremely long  duration of the school-to-work transition in Italy, which has been further prolonged by the COVID-19 pandemic. On overage, an individual in the 20-24 age group in education was expected to find  a permanent job for the first time after 8.63 years or a temporary job after 3.72 years before the pandemic (i.e., in the third quarter of 2019). Within the same age category, females would take 11.70 years and 4.08 years to find a permanent or temporary job, respectively. Residents in the South would take 14.92 years and 4.23 years. The pandemic has worsened all these figures, raising the expected duration of the school-to-work transition to a permanent job  to 11.15 years or a temporary job  to 4.16 years, affecting disproportionately more males and non Italian citizens.

Our analysis also identifies the categories of individuals who are at risk of carrying longer lasting consequences. Most categories of individuals experienced higher probability to lose a job, to become unemployed, to enter into the NLFET (Neither in the Labour Force nor in Education or Training) state  or to join the furlough scheme in the second quarter of 2020. However, in the third quarter of 2020 these probabilities fell back for most categories of individuals, but not for all. In particular, individuals in the 25-29 age category were less employed and populated more the NLFET state, particularly females and non Italian citizens.
Among individuals in the 20-29 age category, we also find marginal evidence  of an increased return to schooling, which is likely to be ascribable to the worse labour market conditions.
Overall, the consensus about  young people in Europe carrying the largest burden of the COVID-19 pandemic \citep{european2020living} is supported by our results which show evidence of a persistent negative effect on the already slow transition from school-to-work, as well as of a remarkable sustained exit of females and non Italian citizens from the labour force in the second and third quarters of 2020.
These findings further strengthen the urgency of policies to support young people in their transition from education to employment and in their opportunity to participate to the labour market.
	
\newpage
	
\bibliographystyle{chicago}
\bibliography{references}
	
\newpage
	
\appendix
\begin{Large}
\textbf{Appendix}
\end{Large}

\section{First passage time and expected first passage time \label{app:firstpassagetime}}

The distribution of times for which a random process arrives for the first time at state $j$ starting from state $i$ is called \textit{the first passage time} from state $i$ to state $j$. The expected time of going from state $i$ to state $j$ for the first time is called \textit{expected first passage time} from state $i$ to state $j$ \citep[p. 818]{lieberman2001introduction}.

The first passage times are random variables, whose probability distributions depend upon the transition probabilities $p_{ij}$. In particular, let $f_{ij}^{(n)}$ denote the probability that the first passage time from state $i$ to $j$ is equal to $n$. For $n>1$, this first passage time is $n$ if 1) the first transition is from state $i$ to some state $k \; (k \neq j)$; and 2) then the first passage time from state $k$ to state $j$ is $n - 1$.
Therefore, these probabilities satisfy the following recursive relationships:
\begin{eqnarray}
f_{ij}^{(1)} &=& p_{ij}; \\
f_{ij}^{(2)} &=& \sum^{K}_{k \neq j} p_{ik} f_{kj}^{(1)};\\
&...& \\
f_{ij}^{(n)} &=& \sum^{K}_{k \neq j} p_{ik} f_{kj}^{(n-1)}.
\label{eq:firstTimePassage}
\end{eqnarray}
Thus, the probability of a first passage time from state $i$ to state $j$ in $n$ steps can be computed recursively from the transition probabilities $p_{ij}$.
Finally, from Eq. (\ref{eq:firstTimePassage}) the expected first passage time from state $i$ to state $j$, denoted by $\mu_{ij}$, can be calculated as:
\begin{equation}
\mu_{ij} = \sum_{n=1}^{\infty} n f_{ij}^{(n)},
\label{eq:expectedFirstTimePassage}
\end{equation}
which is well defined if (\citealp{lieberman2001introduction}):
\begin{equation}
 \sum_{n=1}^{\infty} f_{ij}^{(n)}=1.
 \label{eq:conditionExistenceProbabilityFirstTimePassage}
\end{equation}	

%
%
%
	
\end{document}